\documentclass[useAMS,usenatbib]{mn2e}

\usepackage{graphicx}
\usepackage{amssymb}
\usepackage{subfigure}

\title{The Origin of Segue 1} \author[M. Niederste-Ostholt et al.]
{M.Niederste-Ostholt$^{1}$\thanks{E-mail:mno@ast.cam.ac.uk},
  V. Belokurov$^{1}$, N.W. Evans$^{1}$, G. Gilmore$^{1}$,
  R.F.G. Wyse$^{2}$, \newauthor J.E. Norris$^{3}$\\
  $^{1}$Institute of Astronomy, Madingley Rd,
  Cambridge, CB3 0HA\\
  $^{2}$Johns Hopkins University, Department of Physics and Astronomy,
  3900 North Charles Street, Baltimore, MD 21218, USA\\
  $^{3}$Research School of Astronomy and Astrophysics, Australian
  National University, Mount Stromlo Observatory, Cotter Road,
  \\Weston, ACT 2611, Australia}

\begin{document}

\date{May 2009}

\voffset-.6in

\pagerange{\pageref{firstpage}--\pageref{lastpage}} \pubyear{2009}

\maketitle

\label{firstpage}

\begin{abstract}
  We apply the optimal filter technique to Sloan Digital Sky Survey
  photometry around Segue 1 and find that the outer parts of the
  cluster are distorted. There is strong evidence for $\sim 1^\circ$
  elongations of extra-tidal stars, extending both eastwards and
  southwestwards of the cluster. The extensions have similar
  differential Hess diagrams to Segue 1 and a Kolmogorov-Smirnov test
  suggests a high probability that both come from the same parent
  distribution.  The location of Segue 1 is close to crossings of the
  tidal wraps of the Sagittarius stream. By extracting blue horizontal
  branch stars from Sloan's spectral database, two kinematic features
  are isolated and identified with different wraps of the Sagittarius
  stream. We show that Segue 1 is moving with a velocity that is close
  to one of the wraps.  At this location, we estimate that there are
  enough Sagittarius stars, indistinguishable from Segue 1 stars, to
  inflate the velocity dispersion and hence the mass-to-light ratio.
  All the available evidence is consistent with the interpretation
  that Segue 1 is a star cluster, originally from the Sagittarius
  galaxy, and now dissolving in the Milky Way.
\end{abstract}


\section{Introduction}

Segue 1 was discovered by~\citet{Be07} as an overdensity of resolved
stars in imaging data from the Sloan Digital Sky Survey (SDSS).  It is
located at equatorial coordinates $\alpha_{2000} \approx152^{\circ},\;
\delta_{2000} \approx 16^{\circ}$ and has a heliocentric distance of
$23\pm 2 $ kpc. This corresponds to Galactic coordinates $\ell =
220.5^\circ$, $b= 50.4^\circ$ and a Galactocentric distance of $\sim
28$ kpc. \cite{Be07} suggested that Segue 1 was an extended globular
cluster, possibly associated with the Sagittarius stream.  With its
initially determined half-light radius of approximately $30$ pc, it
would be amongst the largest Milky Way globular clusters such as
Palomar 5. The object is also unusually faint ($M_V\approx-3$) for
its size.

Segue 1 has some points in common with two other discoveries made with
SDSS -- namely Willman 1~\citep{Wi05} and Bootes~II~\citep{Wa07}.  All
three objects have similar absolute magnitudes and half-light radii,
intermediate between the dwarf spheroidal galaxies, which are dark
matter dominated, and the globular clusters, which show no evidence
for dark matter (see Figure $8$ in ~\citep{Wa08}). Determining the true 
nature of these three objects is important as it may shed light on the important 
question of the size of the smallest dark matter haloes in which baryons 
collapsed to form galaxies.

The interpretation of Segue 1 as a globular cluster has recently been
contested by~\citet{Ge09}. Using Keck/DEIMOS spectroscopy, they
measured the radial velocities of 24 stars in Segue 1 with a mean
heliocentric velocity of $\sim 206$ kms${}^{-1}$ and a velocity
dispersion of $4.2 \pm 1.2$ kms$^{-1}$, leading to claims that Segue 1
is a dwarf galaxy rather than a globular cluster.  Assuming these
stars are gravitationally bound to Segue 1 and are in dynamical
equilibrium, then the implied mass-to-light ratio is $\sim
1200$~\citep{Ge09}, which would make Segue 1 the most dark matter
dominated galaxy detected to date.  However, both assumptions are
questionable. If Segue 1 is a globular cluster that is undergoing
tidal disruption, then extra-tidal stars may not be so easy to
distinguish from gravitationally bound members and the hypothesis of
dynamical equilibrium may be a poor one. More seriously, if Segue 1 is
immersed in the Sagittarius stream, then contamination of any sample
of Segue 1 stars by stream stars may be hard to avoid.

In this paper, we analyze SDSS and auxiliary Canada-France-Hawaii
Telescope (CFHT) photometry of Segue 1. We use the optimal filter
techniques pioneered by \citet{Od03} to identify tidal features in \S
2 and \S 3. A stellar population embedded deep in a massive halo would
not be expected to show visible signs of battering by the Galactic
tides. We compute Segue 1's structural parameters in \S 4. Finally, in
\S 5, we extract from the SDSS spectral database blue horizontal
branch stars and used them to identify the kinematical signal from the
Sagittarius stream.  We show that contamination of kinematically
selected Segue 1 stars by Sagittarius stream stars is a serious
problem, and can lead to artificially inflated velocity dispersions.

\begin{figure*}
	\centering
	\subfigure[]{
	\includegraphics[width=0.24 \textwidth]{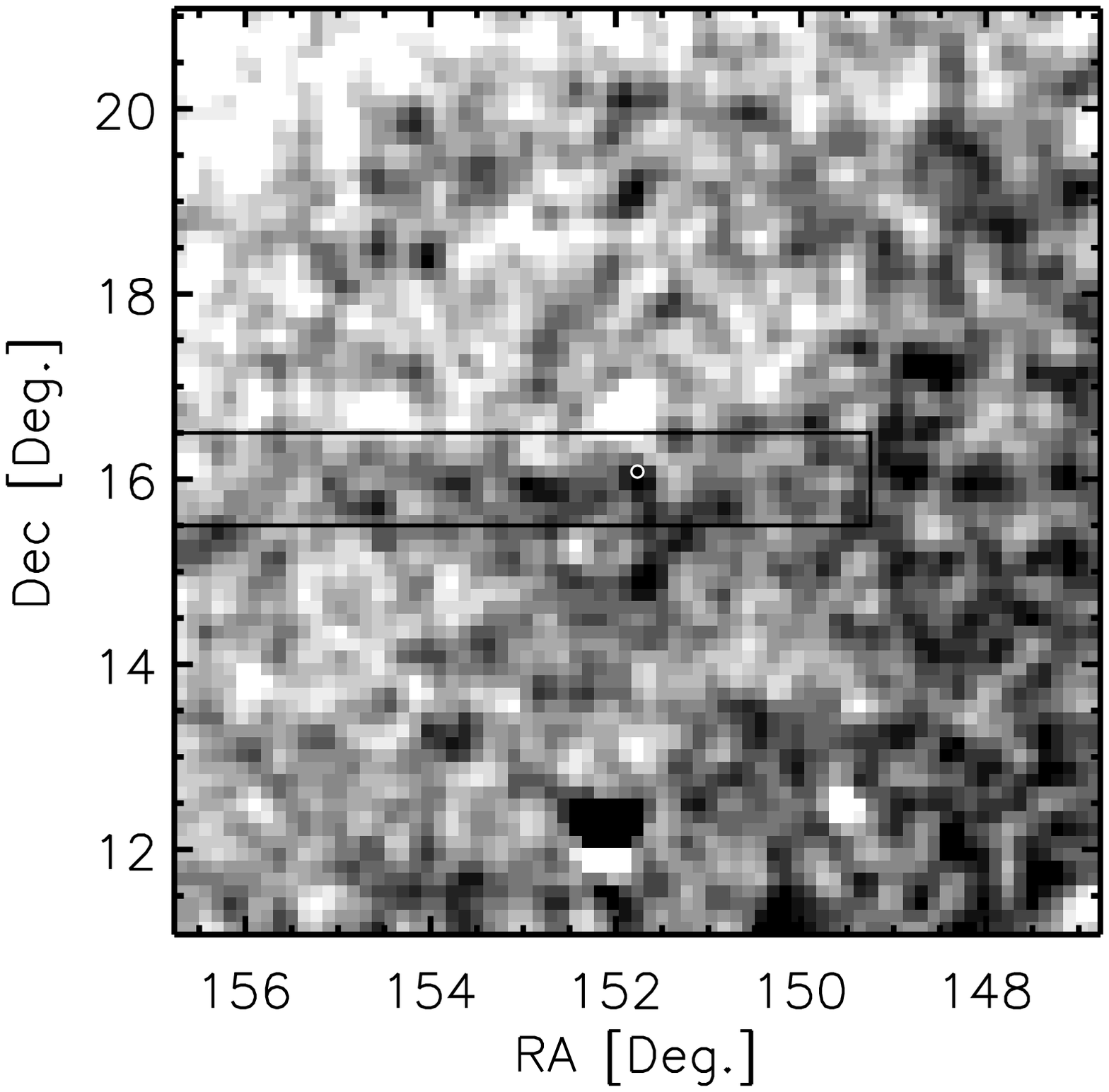}}
	\subfigure[]{
	\includegraphics[width=0.24 \textwidth]{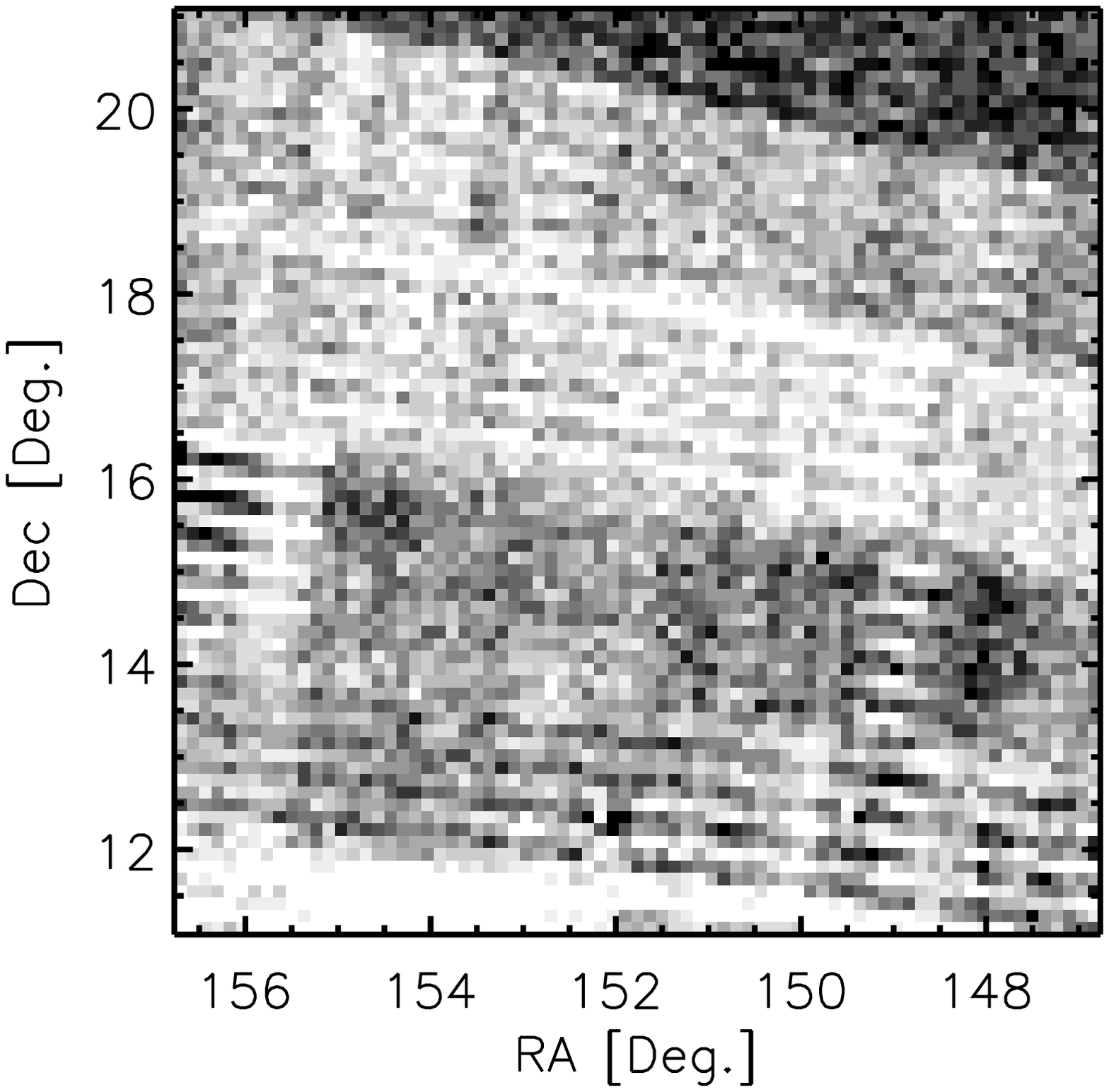}}
	\subfigure[]{
	\includegraphics[width=0.24 \textwidth]{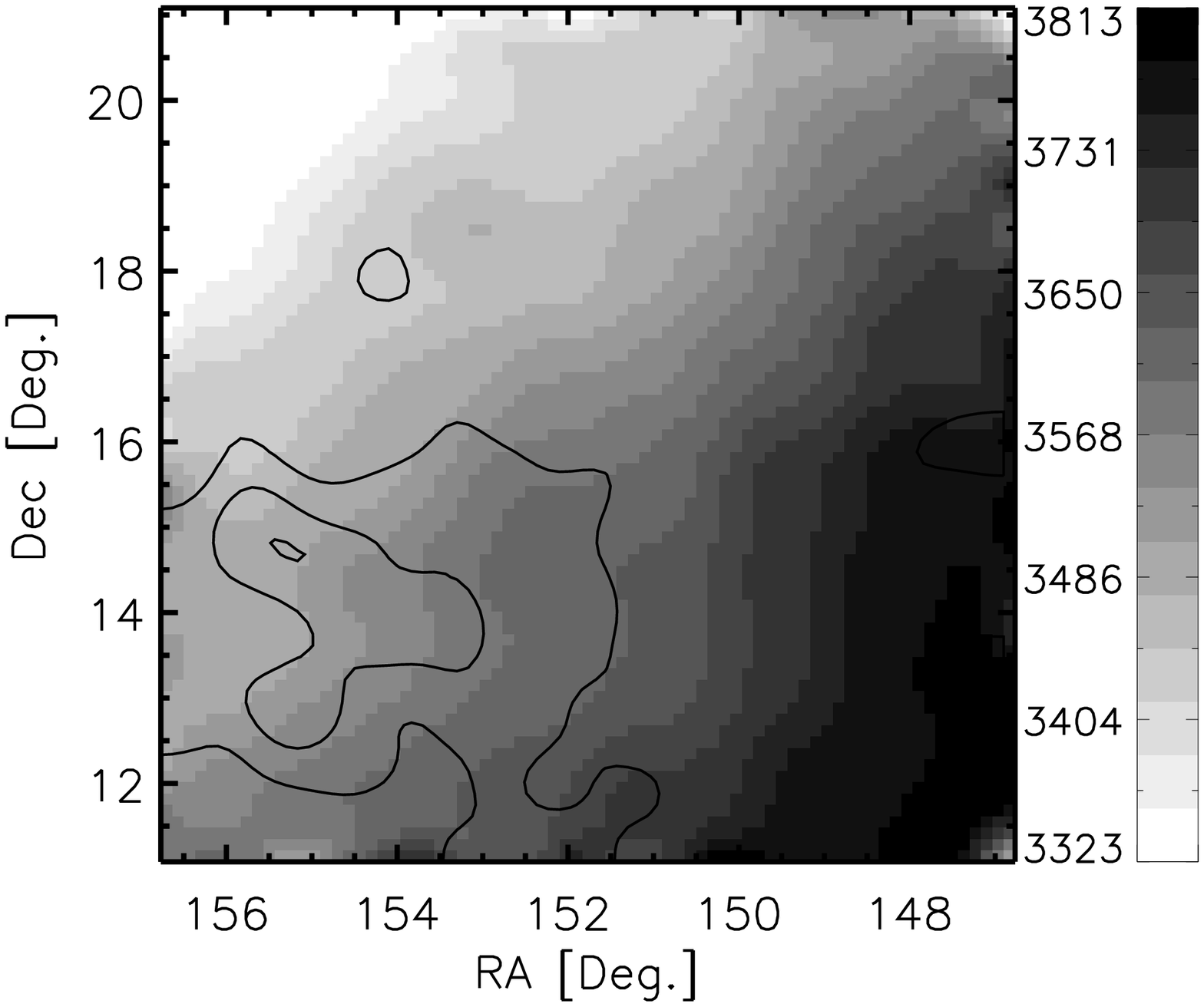}}
	\subfigure[]{
	\includegraphics[width=0.24 \textwidth]{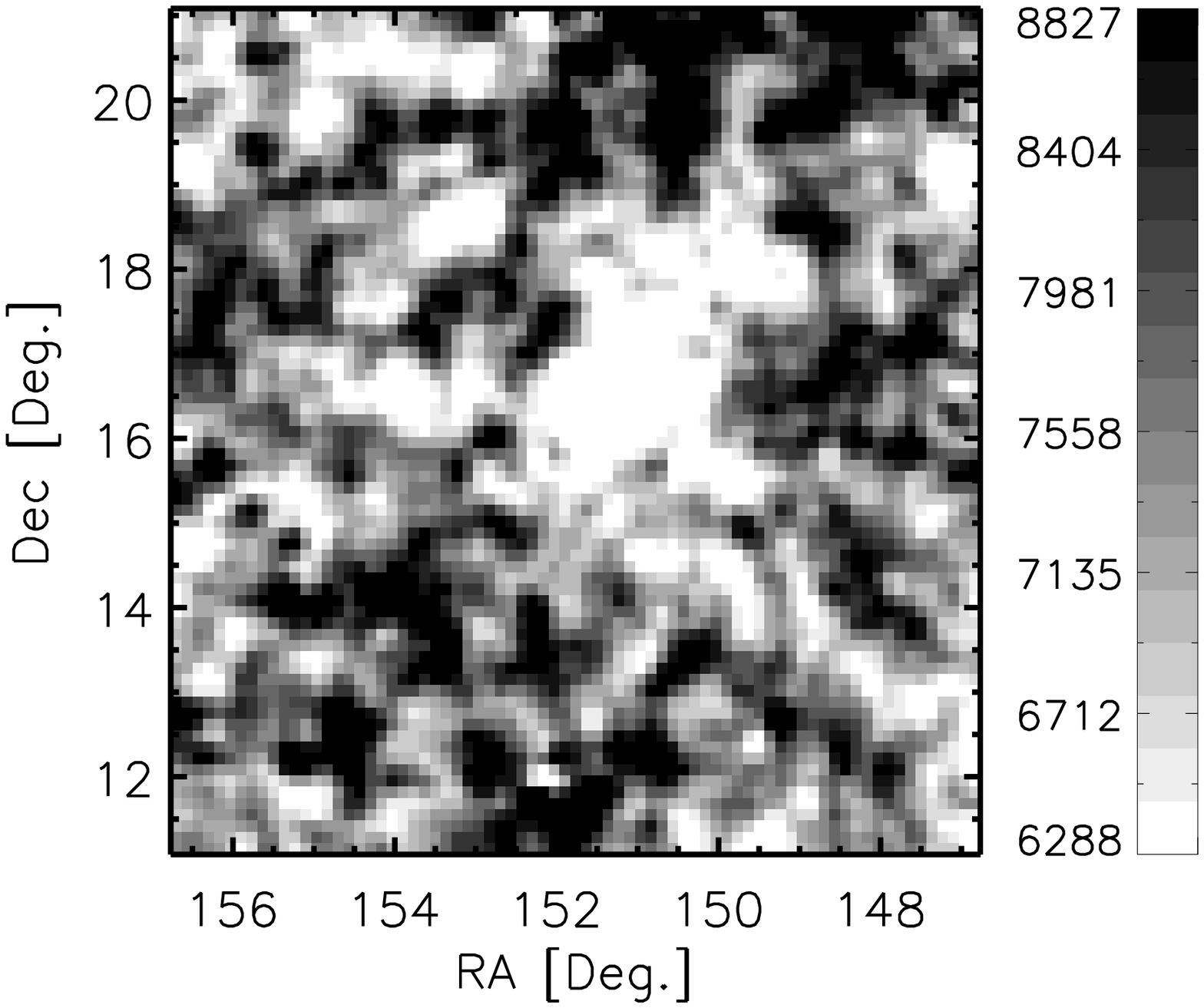}}
      \caption{Panel (a) shows the distribution of all stars with
        clean SDSS photometry with $14<r<22$ in a $10^{\circ} \times
        10^\circ$ square centered on Segue 1 (dark areas are high
        density). The overdensity at $\alpha \approx 152.2^\circ,
        \delta \approx12.5^\circ$ is the Leo I dwarf galaxy.  Also
        shown as a black rectangle is the CFHT footprint, while the
        white circle marks the location of Segue 1. Panel (b) shows
        the density of stars which have been removed from the sample
        using our magnitude cut. The visible bands correspond to the
        SDSS scan directions. Panel (c) shows our estimate of the
        field stars (background and foreground) with extinction
        contours overplotted. The sidebar gives the number of stars
        per square degree. As will become clear later, there does not
        seem to be any correlation between high extinction and any
        extra-tidal features. Panel (d) shows the distribution of
        galaxies from the SDSS data in our field of view. The sidebar
        shows number of galaxies per square degree.}
	\label{fig:density}
\end{figure*}
\begin{figure*}
	\centering
	\subfigure[]{
	\includegraphics[width=0.3 \textwidth]{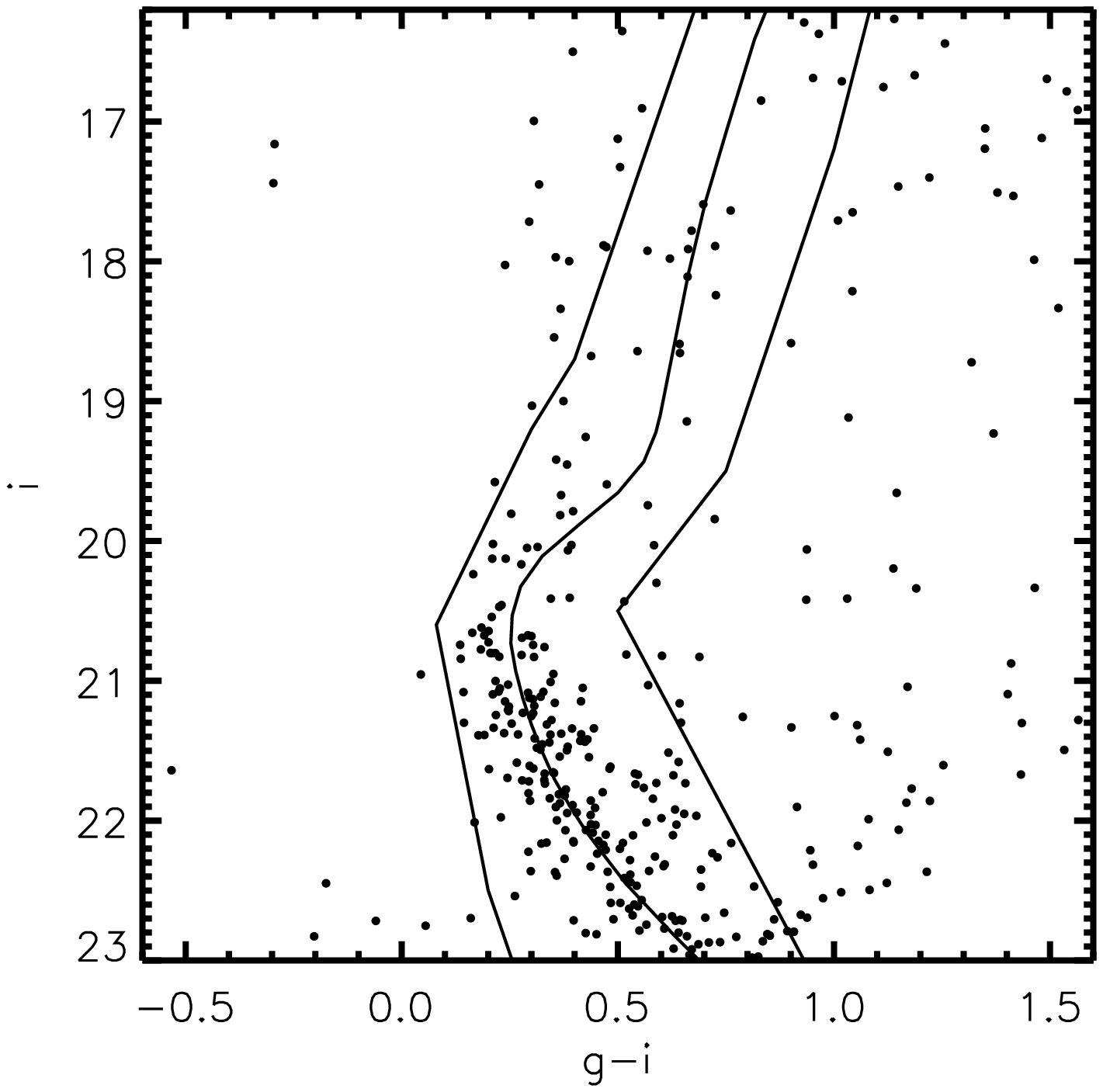}}
	\subfigure[]{
	\includegraphics[width=0.3 \textwidth]{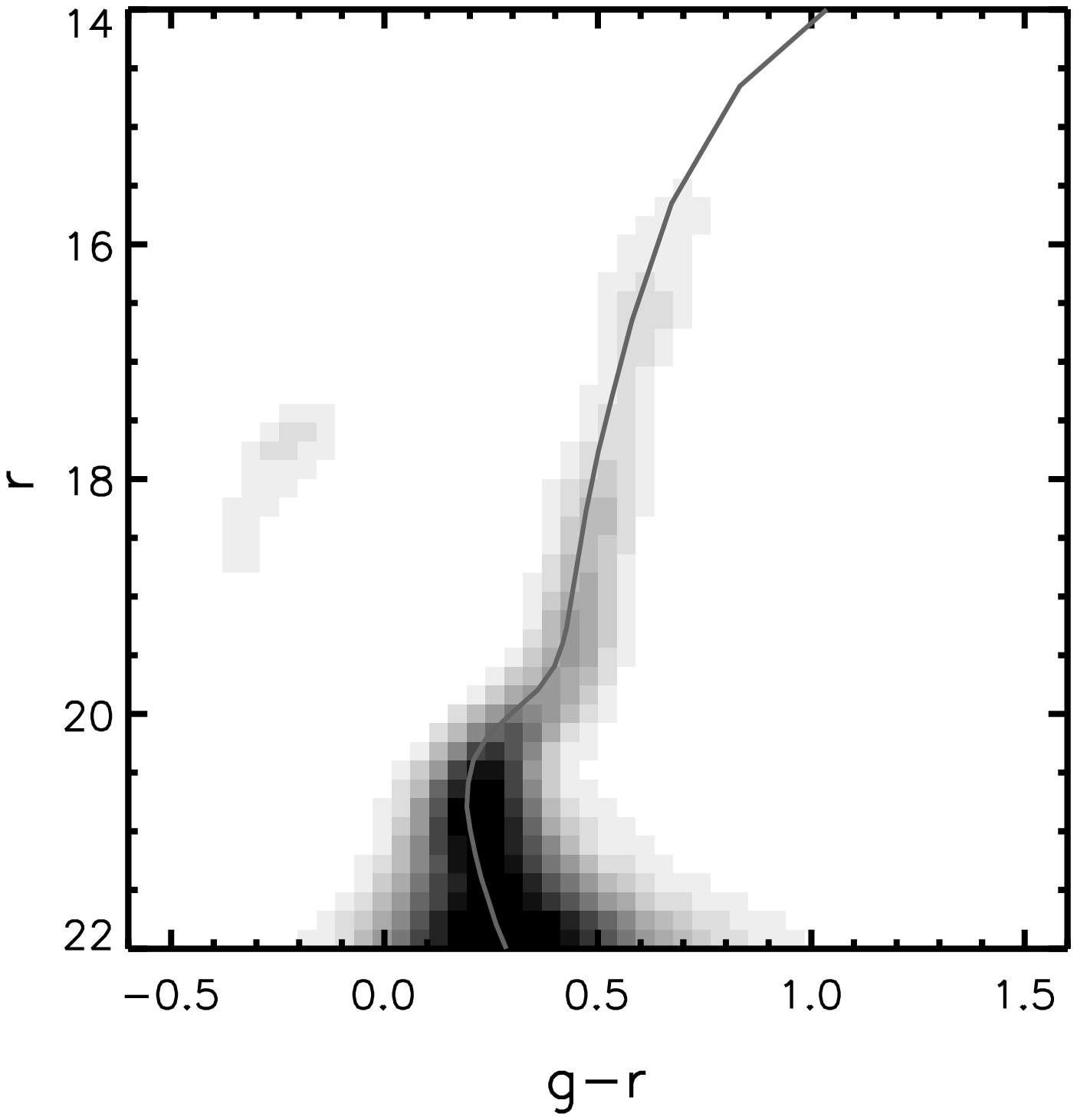}}
	\subfigure[]{
	\includegraphics[width=0.3 \textwidth]{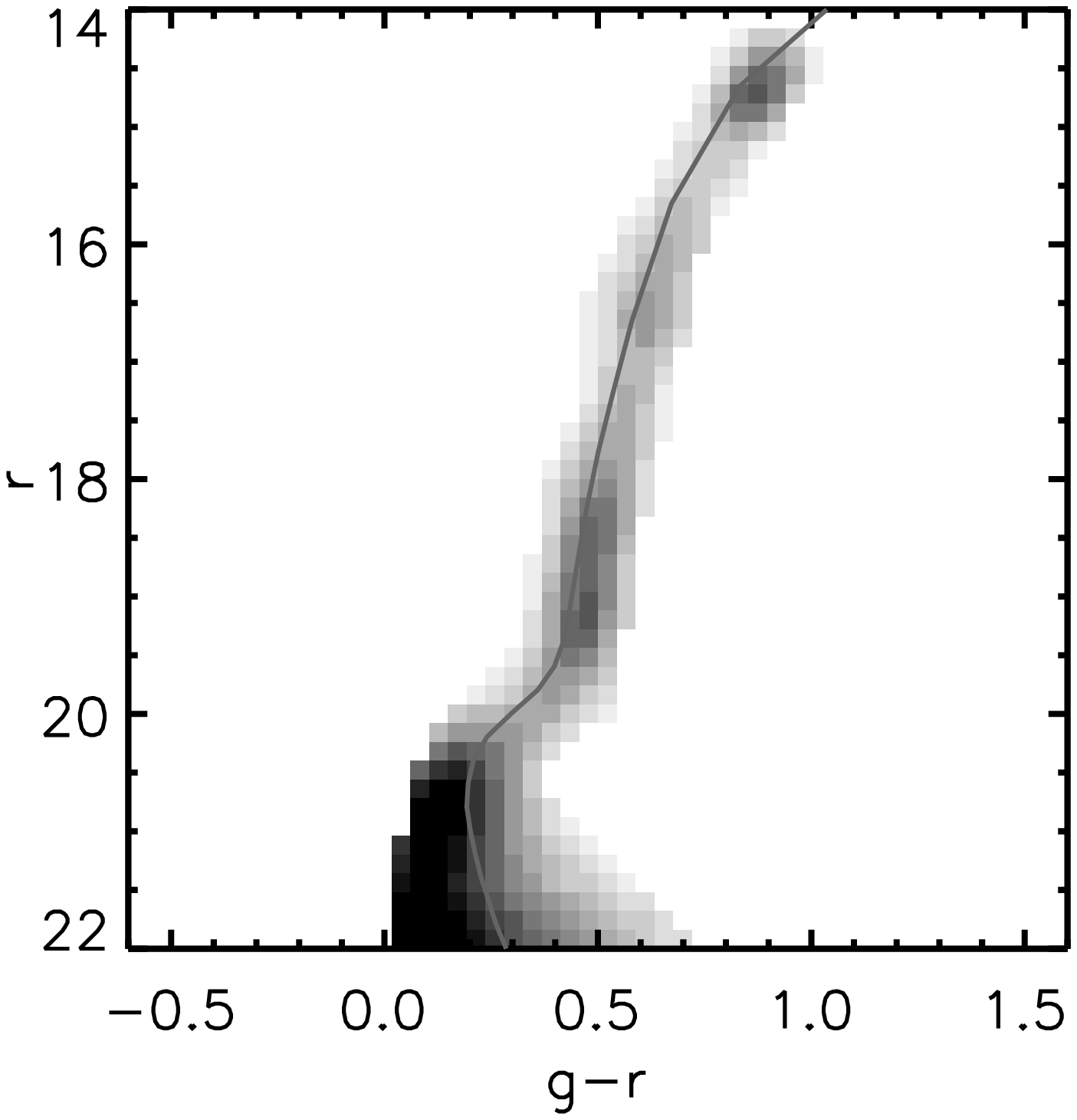}}
      \caption{Panel (a) shows a scatter plot of stars within
        $0.12^\circ$ of Segue 1's center, as seen in the CFHT
        data. The M92 ridgeline is plotted and it is in very good
        agreement with the data, which suggests that we use the more
        complete data on M92 to approximate the CMD of Segue 1. The
        mask overplotted is used to select possible members. Stars
        outside the mask are highly unlikely to be cluster members and
        their weight is zero.  Panel (b) shows the Hess diagram of
        M92. A spread has been added to the data in order to more
        realistically reproduce the SDSS photometry. The spread
        corresponds to the difference between the uncertainty inherent
        in the M92 data and the uncertainty found in SDSS at the
        corresponding magnitudes. Panel (c) shows the ratio of the M92
        Hess diagram and the field Hess diagram. This ratio is used as
        the weights distribution in the optimal filter technique. }
 	\label{fig:cm_m92}
\end{figure*}
\begin{figure*}
	\subfigure[]{
	\includegraphics[width=0.4 \textwidth]{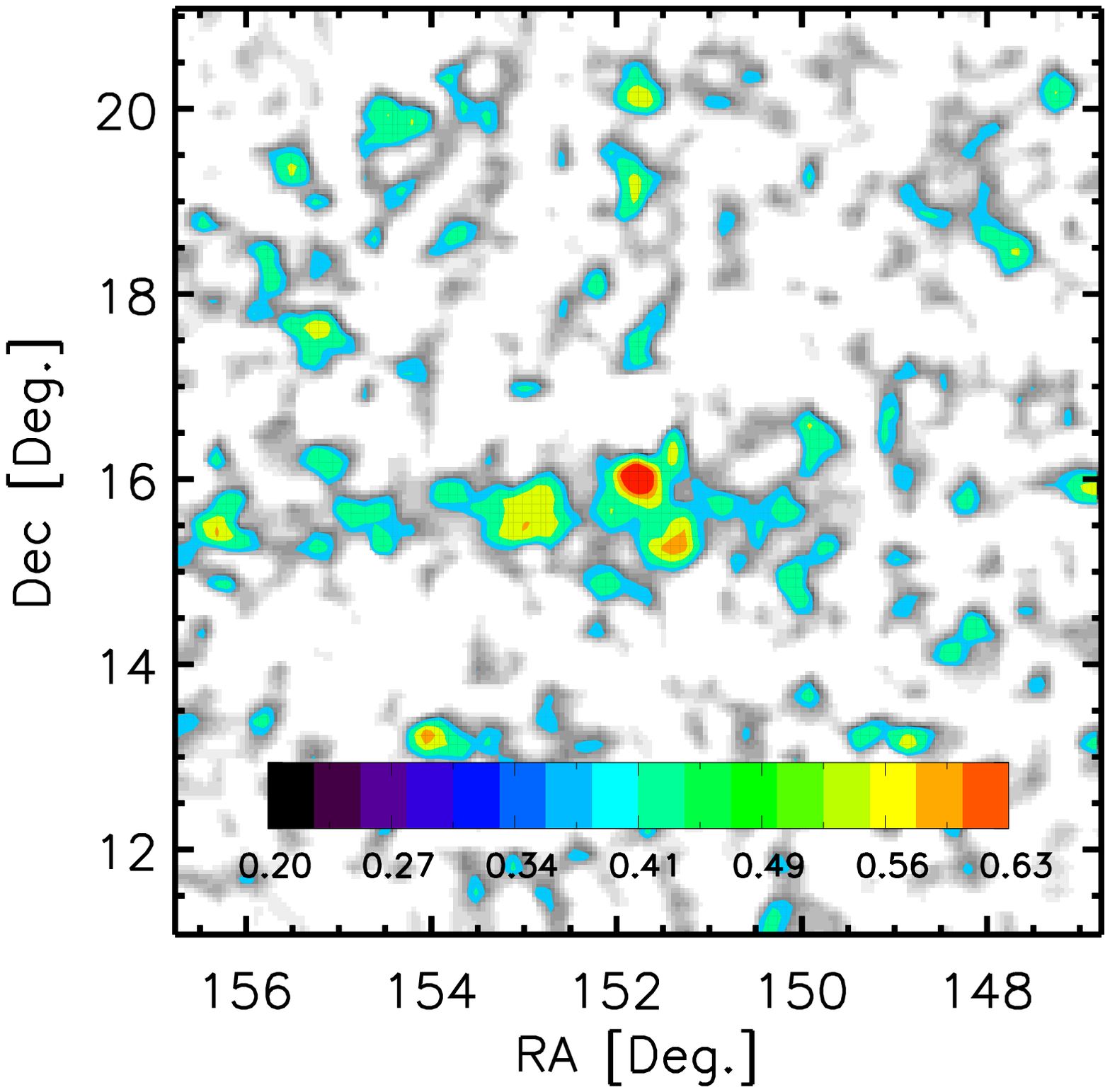}}
	\subfigure[]{
	\includegraphics[width=0.4 \textwidth]{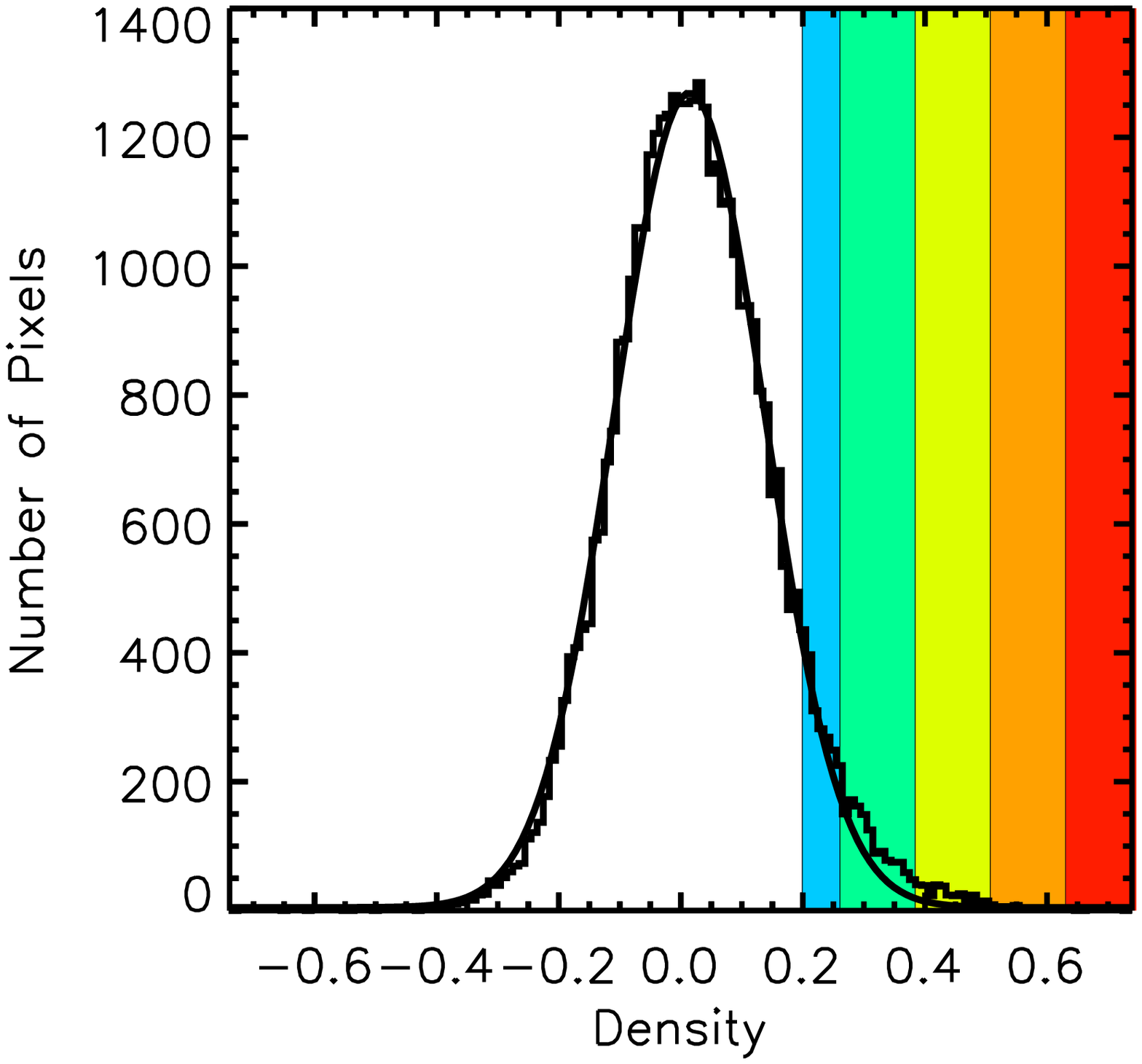}}
      \caption{Panel (a) shows the weighted and background corrected
        Segue 1 star density with a structure extending outwards from
        Segue 1. The density is normalized to the average pixel value 
        within $0.1^{\circ}$ of the center of Segue 1. The contours
        show 1.5,2,3,4 and 5 $\sigma$ levels above average density, 
        whilst the colour within the contours can be converted into 
        normalized pixel density by the key. Panel (b) shows the 
        distribution of pixel values for panel (a) with a fitted
        Gaussian. The vertical colour bars indicate the 1.5,2,
        3, 4, and 5$\sigma$ levels as determined from the Gaussian
        distribution. The Gaussian fits the distribution well, but
        there is an excess at the positive end due to the presence of
        Segue 1. }
	\label{fig:optfilter}
\end{figure*}

\section{Data}

The Sloan Digital Sky Survey (SDSS; York et al. 2000) is an imaging
and spectroscopic survey that covers one-quarter of the celestial
sphere. The SDSS data are described in the data release papers
(Adelmann-McCarthy 2008 for the sixth release, DR 6) and documented at
``http://www.sdss.org.''

We select stars in a $10^{\circ} \times 10^{\circ}$ box centered on
Segue 1 from SDSS DR 6 data, using the SDSS clean photometry flags and
a magnitude cut ($r>22$) to remove data reduction artifacts from the
field. In addition, we correct for extinction using the maps of
\cite{Sc98} and apply the UberCal correction as described in
\citet{Pa07}. Figure \ref{fig:density}a shows the density of all stars
(fainter than $r=14$) selected by dividing the area in $75 \times 75$
pixels and smoothing the field with a Gaussian kernel FWHM of 2 pixels
(this is applied to all density plots in this work). Segue 1 is too
faint to stand out in this figure. Figure \ref{fig:density}b shows a
density plot of stars that we have removed from the sample using our
magnitude cut.  Structural artifacts due to problems in data reduction
are clearly visible and correspond to the SDSS scan patterns on the
sky.  Figure \ref{fig:density}c shows our estimate of the field star
density with extinction contours overplotted. The estimate is
determined by first replacing Segue 1 within $1^{\circ}$ with a
representative patch of the field star distribution taken at
$\alpha=149^\circ,\delta=19^\circ$. By a similar cloning method, we
remove the other obvious overdensity in the field, the Leo I dwarf
galaxy ($\alpha \approx 152.2^\circ, \delta \approx12.5^\circ$, at a
Galactocentric distance of 250 kpc). We then compute the density and
smooth the resulting distribution with a Gaussian kernel with a FWHM
of 5 pixels and a box-car smoothing over 25 pixels. Finally, since
classification of galaxies and stars especially at faint magnitudes
can be uncertain in the SDSS, it is important to verify that possible
structures are not influenced by such misclassifications. Figure
\ref{fig:density}d shows that there are no obvious galaxy
overdensities in the immediate vicinity of Segue 1. There does seem to
be an underdensity, however, this would not generate spurious tidal
structures near Segue 1 except in the highly unlikely event that
almost all galaxies were misclassified as stars.

The main dataset used in this paper is from the SDSS. However, we also
have access to a set of MegaCam pointings taken on the
Canada-France-Hawaii Telescope (CFHT) on 17-25th January 2007.  Data
were taken in $g$ and $i$ filters, typically three 200 s exposures
using the measured zero-points for MegaCam. The MegaCam $g$ and $i$
magnitudes are converted to SDSS magnitudes by matching the stars in
the CFHT data to the SDSS observations. We find 21952 matches between
the two data sets. We then plot the difference in magnitudes in the
two systems versus CFHT color to find the relations
\begin{eqnarray}
g_{\rm SDSS}-g_{\rm CFHT} &=& -0.22+0.21(g_{\rm CFHT}-i_{\rm
  CFHT}) \nonumber \\
& & -0.05(g_{\rm CFHT}-i_{\rm CFHT})^2\nonumber\\
i_{\rm SDSS}-i_{\rm CFHT} &=&-0.09+0.01(g_{\rm CFHT}-i_{\rm
  CFHT})\\
& & +0.01(g_{\rm CFHT}-i_{\rm CFHT})^2\nonumber
\end{eqnarray}
There are 10 MegaCam fields each covering 1 deg$^2$, straddling Segue
1 at fixed declination. A rectangle bounding the 10 fields is shown in
Figure~\ref{fig:density}a. Although deeper than the SDSS data by about
a magnitude, the CFHT data cover considerably less area and are not so
useful for diagnosing tidal features. Nonetheless, it is the best
dataset for estimating the surface brightness profile and hence the
structural parameters of Segue 1.

\begin{table}
\centering
\caption{K-S tests comparing the Hess diagrams of different above
average signal regions and Segue 1. A high K-S probability indicates a
high confidence that the two samples are drawn from the same parent
population. The number of stars is the count of stars inside the CMD mask 
in a given box. The count of Sagittarius stars (i.e. stars falling inside the 
CMD mask but belonging to the Sagittarius stream rather than Segue 1)
is estimated using the methods described in \S 5.1.}
\begin{tabular}{@{}llcccccccc@{}}
\hline
Debris & Area & Number & Number &
K-S \\
Region & [deg$^{2}$] & of Stars & of Sag Stars &
Probability \\
\hline
Segue 1 & 0.045 & 59 & 11 - 25 & \null\\
Box 1 & 1.140 & 1175 & 285 - 627 & 0.93 \\
Box 2 & 1.785 & 1908 & 446 - 982 & 0.99 \\
Box 3 & 0.515 & 607 & 128 - 283 & 0.99 \\
Box 4 & 0.825 & 833 & 206 - 453 & 0.06 \\
Box 5 & 0.350 & 370 & 87 - 193 & 0.55 \\
Box 6 & 0.600 & 592 & 150 - 330 & 0.15 \\
Box 7 & 0.440 & 391 & 110 - 242 & 0.05 \\
\hline
\end{tabular}
\label{tab:ks}
\end{table}

\section{The Optimal Filter Technique}

\subsection{Introduction}

The optimal filter technique works by calculating conditional
probabilities of satellite and foreground membership from densities in
colour-magnitude space, known as Hess diagrams. For each star, the
likelihood (or weight) of Segue 1 membership is simply proportional to
the ratio of the satellite and field Hess diagrams in the relevant
pixel of colour-magnitude space. As the Galactic foreground contains
stars from all possible populations in a large range of distances, the
membership likelihood never reaches certainty.  Therefore, instead of
looking at individual stars, we study the density distribution of
possible members represented by weights summed in pixels on the
celestial sphere.  The method is described in \citet{Od03} and
summarized by the formula
\begin{equation}
  n_C(k)=\frac{\Sigma_j [n(k,j)f_C(j)/f_F(j) - n_F(k,j)f_C(j)/f_F(j)]}
{\Sigma_j f_{C}^{2}(j)/f_F(j)}
\end{equation}
where $n_C(k)$ is the background and foreground corrected, weighted
Segue 1 density in position space; $n(k,j)$ labels the stars that are
in the $j^{th}$ bin in color-magnitude space and in the $k^{th}$ bin
in position space; $f_C$ and $f_F$ represent the color-magnitude Hess
diagrams of Segue 1 and the field respectively (their ratio being the
weight we employ). Finally, $n_F(k,j)$ is the field star density,
which we show in Figure~\ref{fig:density}c, split according to the
bins in colour-magnitude and position space.

We analyzed the color-magnitude distribution of Segue 1 ($f_C$) by
considering all stars that lie within a $0.12$ deg aperture around the
cluster center. The Hess diagram of Segue 1 is shown in Figures 3 and
4 of \citet{Be07}, with the ridgeline of M92 from \citet{Cl05} at the
distance modulus $16.8$ mag overplotted.  We determine the field star
color-magnitude distribution ($f_F$) by considering stars outside of
an aperture of $0.4^\circ$. We find that optimal results are achieved
by subdividing color-magnitude space with 50 by 50 pixels and
smoothing the resulting distribution with a Gaussian kernel with a
FWHM of 3 pixels in both cases. In the optimal filter technique, the
ratio of the two Hess diagrams is used as the weight to determine
whether a star is a member of Segue 1. Guided by the ridgeline, a mask
about the relevant region of the Hess diagram is tightly drawn. The
weight distribution which we derive using this method is rather
choppy, due to the small number of stars found close to the clusters
center ($<100$ stars) and such a distribution seems unphysical.

Using the deeper CFHT data, we find that the agreement between the
Segue 1 CMD and the M92 ridgeline is sufficiently close
(Figure~\ref{fig:cm_m92} a) to approximate the CMD of Segue 1 with
that of M92. Figure~\ref{fig:cm_m92}b shows the M92 Hess diagram, with
a spread added in order to more accurately reflect the SDSS
photometry, as well as the ratio of M92 and field Hess diagrams which
we ultimately use as weights (Figure~\ref{fig:cm_m92}c). It is
noticeably smoother than the weights distribution generated by using
SDSS data alone.

The foreground is composed of the smooth halo population and some
Sagittarius material. While the mean chemical properties of the outer
halo are not very different from M92~\citep{Ca07}, halo stars are of
course much more spread out both spatially and in
metallicity. Sagittarius stars at this location occupy a slightly
brighter and redder part of the CMD, because Sagittarius is more
metal-rich. Nonetheless, there is an overlap between Sagittarius stars
and Segue 1 stars on the CMD. In other words, an M92 mask centered on
Segue 1's sequence will pick up some Sagittarius stars. 


\begin{figure*}
	\vspace{-20 pt}
	\subfigure[]{
	\includegraphics[width=0.23 \textwidth]{optfilter_gr.ps}}
	\subfigure[]{
	\includegraphics[width=0.23 \textwidth]{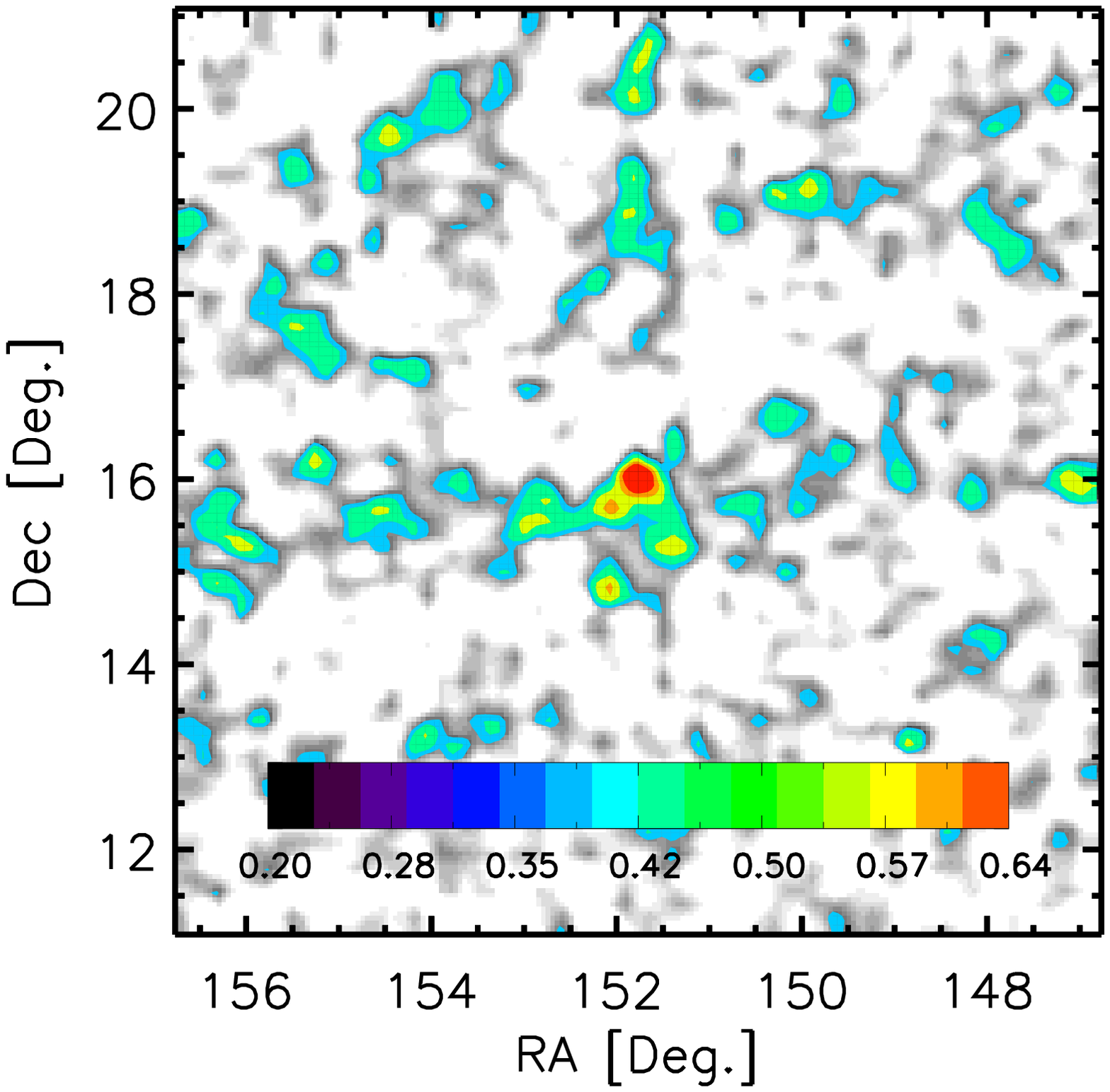}}
	\vspace{-70pt}
	\subfigure[]{
	\includegraphics[width=0.23 \textwidth]{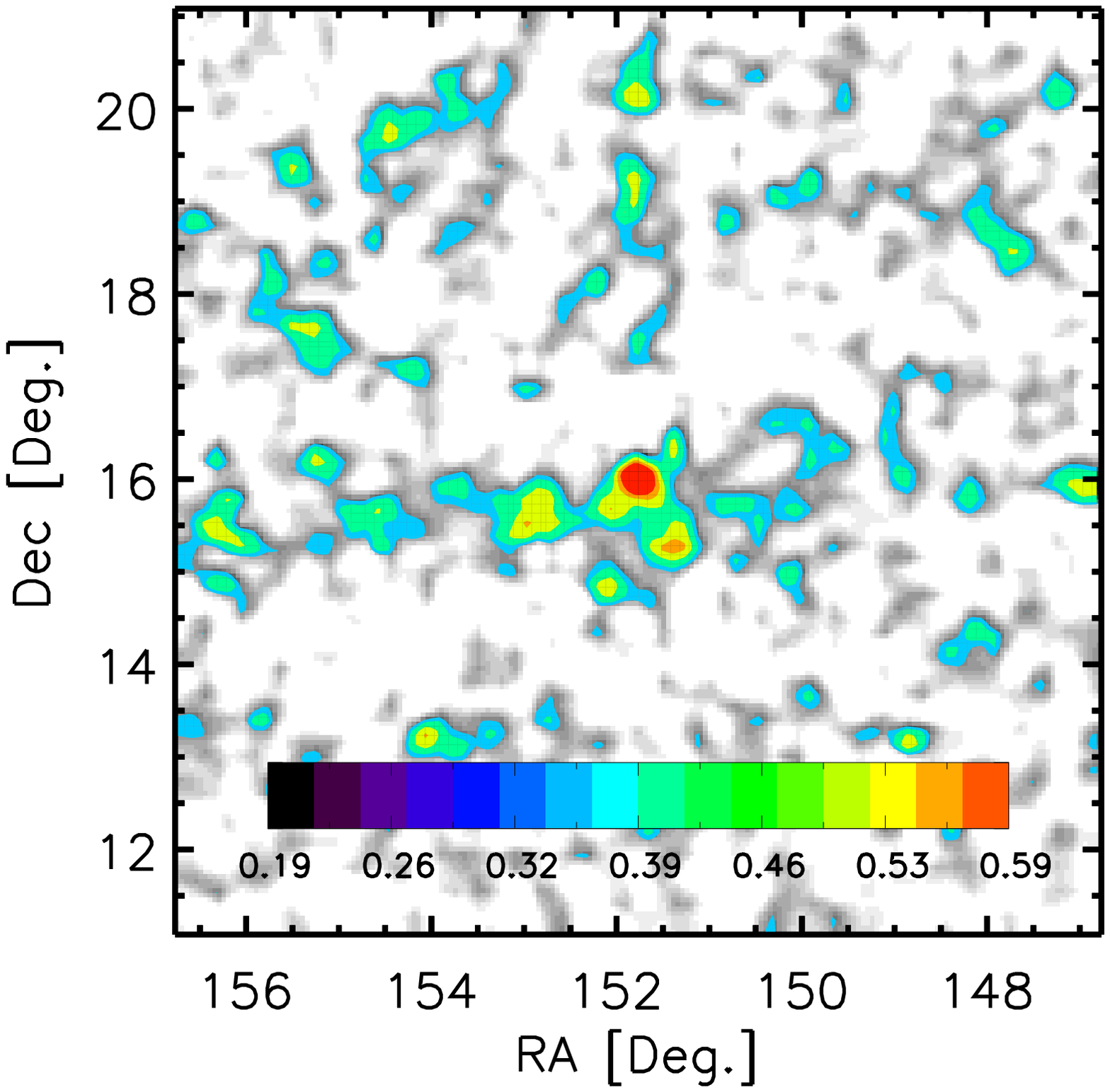}}
	\vspace{-20 pt}
	\subfigure[]{
	\includegraphics[height=0.8 \textwidth]{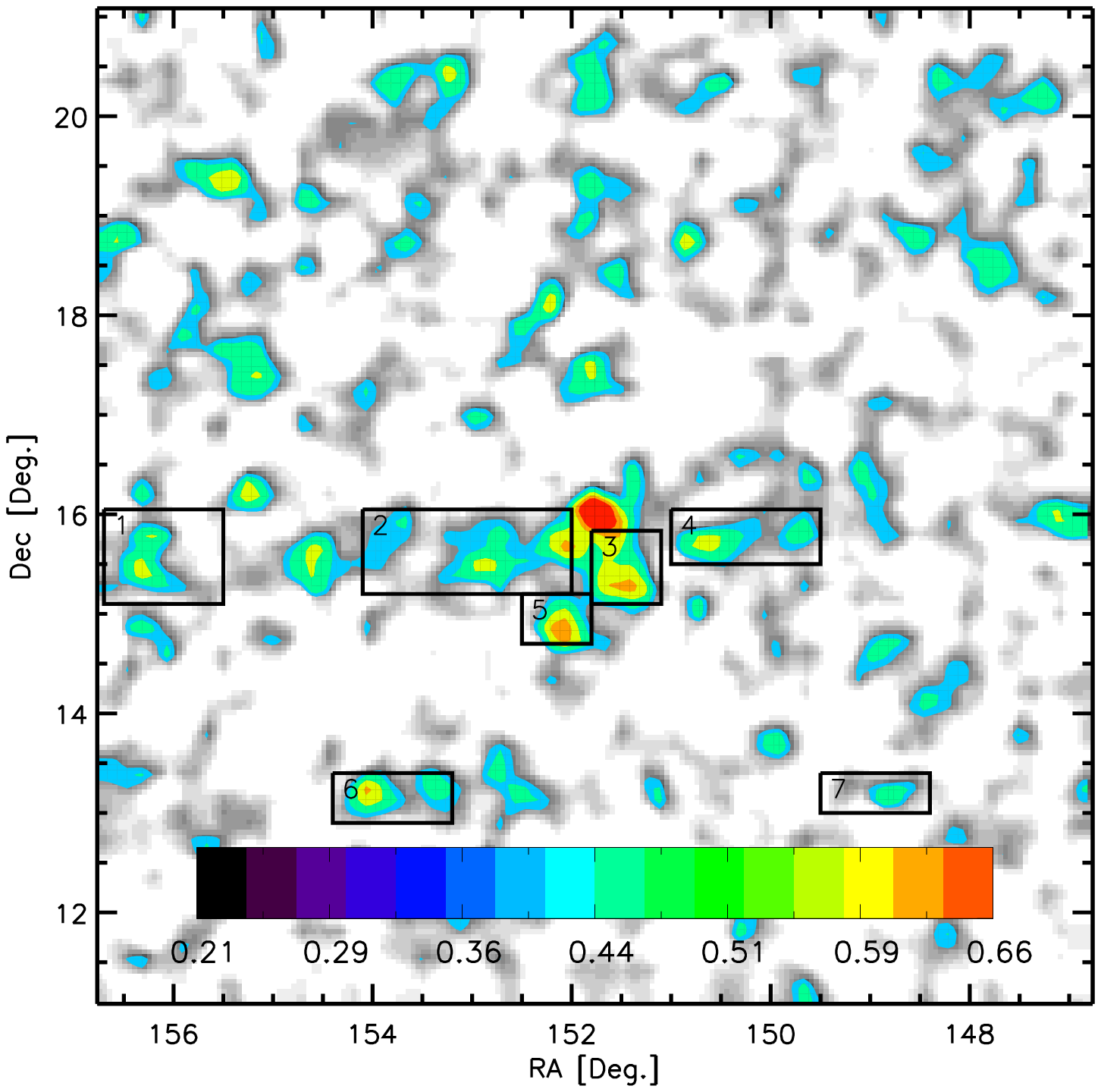}}
	\subfigure[]{
	\includegraphics[width=0.95 \textwidth]{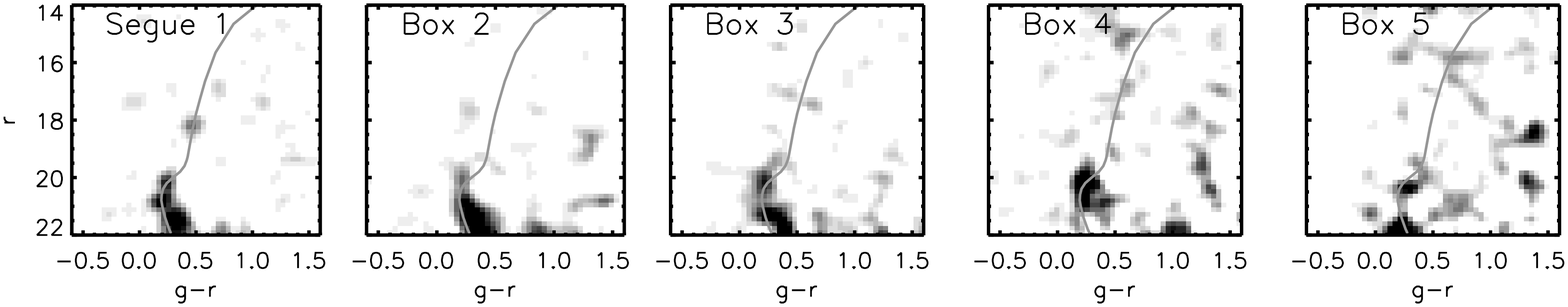}}
      \caption{Panel (a) shows the result of the optimal filter
        technique with weights defined from the $g\!-\!r$ versus $r$
        Hess diagram. Panel (b) shows the results using weights from
        $g\!-\!i$ versus $i$. Panel (c) is the average of the figures
        shown in panels (a) and (b). Panel (d) shows the results of
        the optimal filter technique using weights from the $c_1$
        versus $i$ Hess diagram.  Although there are slight
        differences between our different implementations of the
        optimal filter technique, all four plots support that there
        are extra-tidal features around Segue 1. Panel (e) shows
        differential Hess diagrams of Segue 1 and boxes 2-5
        respectively. Boxes 2 and 3 are very similar to Segue 1
        whereas boxes 4 and 5 are not, even though they have
        reasonably high significance in the optimal filter analysis.}
	\label{fig:optfilters1}
\end{figure*} 

\subsection{Segue 1 as seen by the Optimal Filter}

Using the optimal filter technique with the stellar weights described
above, we find an extended structure which appears to be connected to
Segue 1.  In Figure~\ref{fig:optfilter}a, we show the stellar density
$n_C$. The filled coloured contours indicate the 1.5,2,3,4, and 5
$\sigma$ levels above average density. There is a clear structure
extending in a horizontal band from the center of Segue 1 out to
approximately $\alpha=154^\circ,\delta=15.8^\circ$. It is
approximately $0.5^{\circ}$ wide, and lies along the direction of the
Sagittarius tidal tail at this location. In addition, there appears to
be a structure extending downward from the
cluster. Figure~\ref{fig:optfilter}b shows the distribution of stellar
weight counts in the pixels. We associate the excess of the
distribution above the fitted Gaussian at the positive end with Segue
1 and its tidal debris.

We repeat the optimal filter technique defining the stellar weights
using the $g-i$ versus $i$ Hess diagram as well as the $c_1$ versus
$i$ Hess diagram. Here, $c_1$ is a color index introduced
by~\citet{Od02} and defined by us as:
\begin{equation}
c_1=0.918(g-r)+0.397(r-i)
\end{equation}
This color index is chosen to lie along the one-dimensional
distribution of M92 stars in the $g-r$ versus $r-i$ space. We note
that the stars which we select from the SDSS as being in the center of
Segue 1 do not lie along a one-dimensional locus. We attribute this to
contamination from foreground stars. By creating this color index, the
greatest amount of available data is used and hence the results of the
optimal filter technique based on it should be most robust. In Figure
\ref{fig:optfilters1} we show 4 different views of the optimal filter
technique with weights derived from $g-r$ versus $r$, $g-i$ versus $i$
and $c_1$ versus $i$ Hess diagrams. In addition, we show the average
of the $g-r$ versus $r$ and $g-i$ versus $i$ based optimal filter results.

\subsection{Extra-Tidal Features}

It is essential to verify that the visually identified structures are
actually related to the cluster and not merely the result of chance
alignments with noise within the field. To this, end we analyze the
Hess diagrams of a number of overdensities in the field. 

We investigate 7 areas of overdensity near Segue 1 in more detail. For
each of the boxes shown in Figure~\ref{fig:optfilters1}, we generate
differential Hess diagrams in order to determine if the stellar
populations are similar to that of the core of Segue 1. The lower
panels of Figure~\ref{fig:optfilters1} show some sample results. We
find that boxes 2 and 3 (which represent tidal debris directly
connected to the cluster) have differential Hess diagrams that follow
the theoretical ridgeline of M92 closely. The slight deviation from
the ridgeline at faint magnitudes is attributed to deficiencies in the
SDSS photometry (also visible for example in the Hess diagram of
Belokurov et al. 2007). Boxes 4 and 5 lie close to the cluster and
have a high significance. However, we find that their differential
Hess diagrams do not agree with that of Segue 1. Box 1 is a high
signal area that lies on what seems to be an extended row of
overdensities near Segue 1, but again seems to have a dissimilar
differential Hess diagram. Boxes 6 and 7 represent two overdensity
patches clearly below the cluster (with box 7 only appearing as a
significant overdensity in the $g-r$ and $g-i$ implementations of the
optimal filter technique). Their differential Hess diagrams do not
look like those of Segue 1.

We assess the similarity between the differential Hess diagrams of the
boxes with that of Segue 1 by using the Kolmogorov-Smirnov (K-S)
test. In order to apply the K-S test, we in effect determine a
luminosity function from the $g\!-\!r$ versus $r$ Hess diagrams by
summing pixel rows and combining the counts in adjacent rows
(i.e. doubling the size of the pixels in the magnitude direction). To
reduce the influence of foreground stars on our test, we exclude noise
seen in the Hess diagrams redward of $g-r = 0.8$.  Using only stars in
a narrowly defined mask about the ridgeline proved impractical because
the noise in the Hess diagrams is too great.  We calculate cumulative
luminosity functions and determine the maximum difference between the
cumulative distribution found in the box to that of Segue 1.  The
results of the K-S tests are summarized in Table \ref{tab:ks} with a
high probability suggesting that the two distributions come from the
same parent distribution. The table also gives the number of stars
within each box, together with an estimate of the number of
Sagittarius stream stars (see Section 5.1). We caution that smoothing
the data may have introduced correlations and so the K-S probabilities
may be slightly too large.  We conclude that the overdensities in
boxes 2 and 3 are consistent with being tidal material stripped from
Segue 1, at least as far as the Hess diagrams are concerned.

\begin{figure}
\includegraphics[width=0.5\textwidth]{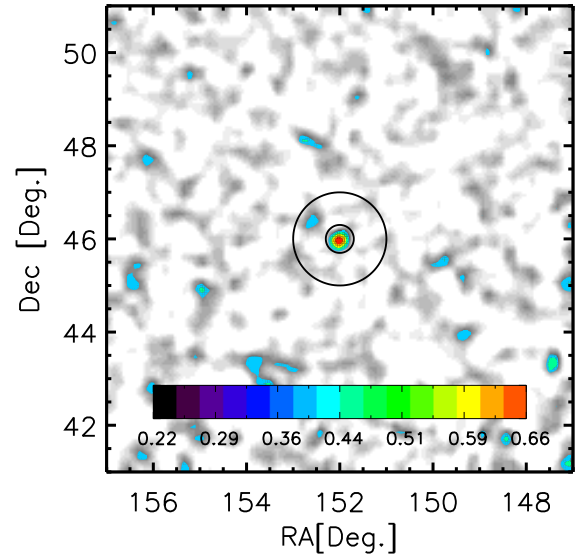}
\caption{The optimal filter technique is applied to input data in
  which a simulated Segue 1 lies in a part of the sky uncontaminated
  by the Sagittarius stream. The field star distribution is much
  smoother and the noise is significantly reduced as compared to
  Figures~\ref{fig:optfilter} and \ref{fig:optfilters1}}.
\label{fig:simul}
\end{figure}

\begin{figure}
	\includegraphics[width=0.5 \textwidth]{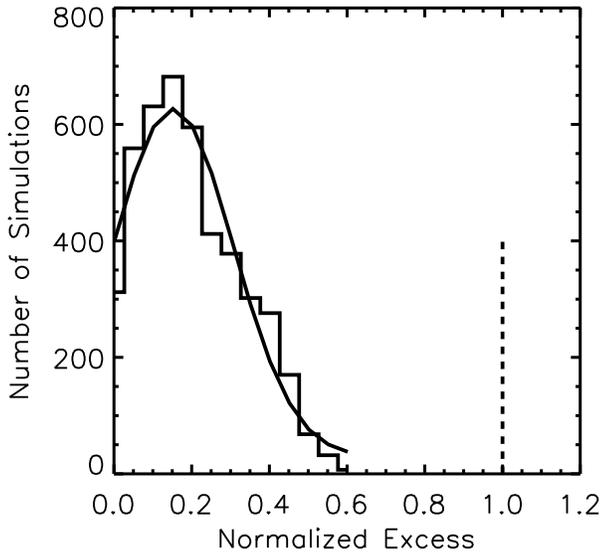}
	\caption{The distribution of excess measured around the
          simulated Segue 1 objects placed randomly in the field of
          view.  The horizontal axis show the excess as a normalized
          count i.e. the ratio of the count around the simulated
          object to the count around Segue 1. The vertical axis
          indicates how many simulations had this excess. The dashed
          line indicates the location of the excess around Segue 1.
          Depending on the Hess diagram used to define the weights in
          the optimal filter technique, the excess around Segue 1 is
          at least a $3 \sigma$ outlier. Here we show the results of
          the simulation with weights defined by the $c_1$ versus $i$
          Hess diagram.}
	\label{fig:simhisto}
\end{figure}

Finally, we also investigate whether or not the extra-tidal features
visible in Figure~\ref{fig:optfilters1} are the result of chance
alignments with the noise in our field of view. To this end, we place
4424 simulated Segue 1 like objects in the field.  These are created
by randomly sampling the Plummer profile fit to the CFHT data and the
luminosity function generated from M92.  We perform the optimal filter
analysis and measure the stellar excess around Segue 1 and the
simulated object by counting the stellar weights in a circular annulus
between 0.3 and 1 degree from the cluster center, as indicated in
Figure~\ref{fig:simul}. We count only those pixels that are at least
$1.5 \sigma$ above the average density. The excess measured around
Segue 1 is a significant outlier (at least 3$\sigma$) compared to the
excess measured about the simulated objects, as shown in Figure
\ref{fig:simhisto}.

Note that the noisy background in Figures~\ref{fig:optfilter} and
\ref{fig:optfilters1} seems to be a consequence of the location of
Segue 1 in the Sagittarius stream. Suppose we place a simulated Segue
1 in a patch of the sky at equatorial coordinates $\alpha =
152^\circ,\delta=46^\circ$, which is well away from the
Stream. Applying the optimal filter technique yields the results shown
in Figure~\ref{fig:simul}. The object is cleanly and easily recovered
and the background has very little prominent substructure. This
suggests that the ``noise'' in our earlier Figures is probably real
substructure in the Sagittarius stream, although further
investigations are need to confirm this.

To conclude, we have demonstrated that there are structures (e.g.,
boxes 2 and 3) seemingly connected to Segue 1 and which have very
similar differential Hess diagrams to Segue 1. The structures are
unlikely to be caused by chance alignment with noise generated by the
Sagittarius stream. It is very natural to ascribe them to tidal
features.  Note that Martin et al (2008) have argued that, even with
the smoothing that tends to obliterate small-scale structure,
realizations of purely spheroidal models with small numbers of stars
can appear just as distorted as the observed photometric data on some
of the ultra-faint galaxies. However, their calculations do not
directly address the existence of tidal features in the outer parts of
these objects.


\section{Structural Properties of Segue 1}

Using available CFHT data, we are able to get a view of a smaller, but
deeper, field containing Segue 1. Performing the optimal filter
analysis on the data proved difficult since the field is too small to
define a meaningful background. We are only able to identify the
cluster but not any prominent extra-tidal features found using the
SDSS data. We are able to confirm that this is not due to differences
or anomalies in the SDSS and CFHT data by applying the optimal filter
technique to a field the size of the CFHT field of view cut out of our
SDSS data. Doing this, we find a similar result that the cluster can
be isolated, but not any obvious extra-tidal features.

\begin{figure}
	\centering
	\hspace{-40 pt}
	\includegraphics[width=0.55\textwidth]{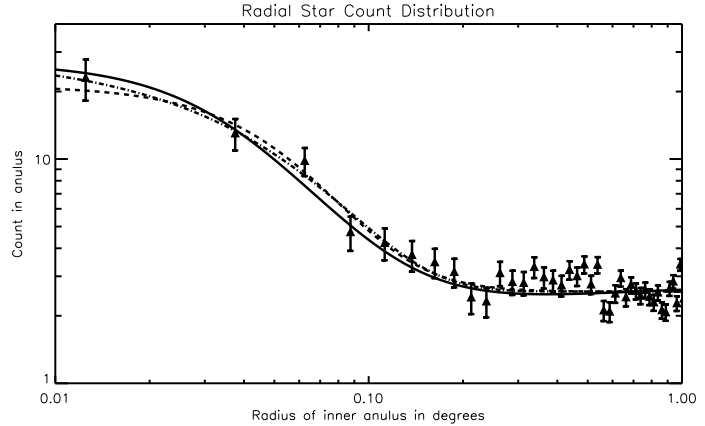}
	\caption{The points show the number of stars in successive
          annuli about the cluster center (annuli width of
          $0.025^{\circ}$) with error bars assuming Poisson
          statistics. We only count stars that fall inside the CMD
          mask. The overplotted lines show King (solid line), Plummer
          (dashed line), and Exponential (dash-dotted line) fits to
          the distribution.  Using the fits we determine a number of
          statistics for Segue 1 which are summarized in Table
          \ref{tab:params}. Note the succession of points $1 \sigma$
          away from the fits in the outer regions of Segue 1, which
          could confirm extra-tidal populations.}
	\label{fig:tanulifit}
\end{figure}

With the CFHT data, we analyze the cluster density profile and
luminosity in greater detail. Figure~\ref{fig:tanulifit} shows
Plummer, King, and exponential profiles fit to the surface density of
Segue 1.  The parameters determined from the fits are summarized in
Table~\ref{tab:params} and compared to the values determined earlier
by Belokurov et al. (2006) and Martin et al. (2008). In the fitting,
only those stars that fall inside the CMD mask of Segue 1 shown in
Figure~\ref{fig:cm_m92} are considered. There is evidence for
extra-tidal populations in the succession of datapoints that lie above
the assumed fit in the outer parts. Even though the deviations from
the fit are always within 1 $\sigma$, there are $\sim 10$ such
datapoints. Within 3 half-light radii, the fit to the models (King,
Plummer or Exponential) is excellent with a $\chi^2$ of 1.1, 0.9 and
0.9 respectively. However, if the datapoints between 3 and 7
half-light radii alone are used, the $\chi^2$ values of the fit are
4.7, 3.6 and 3.7 respectively, emphasising that the outer parts
deviate from the smooth model.

As noted by Martin et al. (2008), determining the luminosity for faint
satellites with few observed member stars is challenging as it is
strongly dependent on the inclusion or exclusion of a small number of
stars that have evolved high up the red giant branch. We determine the
luminosity of Segue 1 in three different ways, each using the CMD
mask: (1) counting the flux within the Plummer half-light radius
determined from our fits and doubling this to get the total
luminosity, (2) determining the flux within the half-light radius and
scaling this up to $10^{\prime\prime}$, (3) counting the flux within
$10^{\prime\prime}$. In each case, we subtract an estimate for the
field star flux and add in a proxy for the missing flux that is lost
due to our magnitude cut-off. The field star flux is determined by
considering stars that fall within our CMD mask and that are in areas
near the cluster but that our optimal filter technique shows to
contain few cluster stars (two boxes defined from
$\alpha=152.2^{\circ}$ to $\alpha=156.3^{\circ}$ and from
$\delta=16.1^{\circ}$ to $\delta=16.6^{\circ}$ and
$\alpha=149.3^{\circ}$ to $\alpha=151.1^{\circ}$ and from
$\delta=16.1^{\circ}$ to $\delta=16.6^{\circ}$). The missing flux is
determined from the amount of flux found below our cutoff in M92.
However, as our magnitude cut-off in the CFHT data is $i = 23$, very
little flux is missed (less than 1\%). The determined values are
summarized in Table \ref{tab:params}.

\begin{table}
\begin{minipage}{0.5\textwidth}
 \centering
  \caption{Structural Parameters of Segue 1. Subscripts P, K and
    E refer to the Plummer, King and Exponential Fits, whilst
    all uncertainties are determined via bootstrapping.}
  \begin{tabular}{@{}llllllllll@{}}
  \hline
 & Be06 & Ma08 & This paper \\
 \hline
$r_{\mathrm c,K}$ & & &$2'.3 \pm 0'.4$ \\
$r_{\mathrm t,K}$&  & & $26'.4 \pm 1'.9$ \\
$r_{\mathrm h,P}$ & $4'.5$ & &$4'.4 \pm 0'.5$\\
$r_{\mathrm h,E}$ & $4'.6$ & $4'.4^{+1.2}_{-0.6}$&$4'.1  \pm 0'.5$ \\
$N_{\star}$ & & $65 \pm 9$ & $83$\\
$M_{tot,V}$  & $-3.0 \pm 0.6$ & $-1.5^{+0.6}_{-0.8} $&
$-2.2\pm0.3^{b}$\\
\null & \null & \null &  $-2.7\pm0.3$ \\
\null & \null & \null &  $-1.6\pm0.3$ \\
 $\mu_V$& & $27.6^{+1.0}_{-0.7}$
\footnote{\footnotesize The definition of surface brightness
   in Martin et al. takes into consideration the ellipticity of the
   cluster, whereas we calculate an effective surface brightness as
   the total flux divided by the area within the plummer half-light
   radius. Using our method and $L_V$ and $N_{\star}$ from Martin et
   al. one would find $\mu_V = 28.6$. } & $27.6 \pm 0.3
 $\footnote{\footnotesize Note that the
    three different values quoted for the absolute magnitude, central
    surface brightness and total luminosity correspond to the three
    different methods described in the main text. }\\
\null & \null & \null & $27.1 \pm 0.3$\\
\null & \null & \null & $28.1 \pm 0.3$\\
 $L_V $($L_{\sun}$) & & $335^{+235}_{-185}$ & $554 \pm 165^{b} $ \\
\null & \null & \null & $960 \pm 286$\\
\null & \null & \null & $364 \pm 147$\\
\hline
\end{tabular}
\label{tab:params}
\end{minipage}

\end{table}

\begin{figure}
\includegraphics[width=\linewidth]{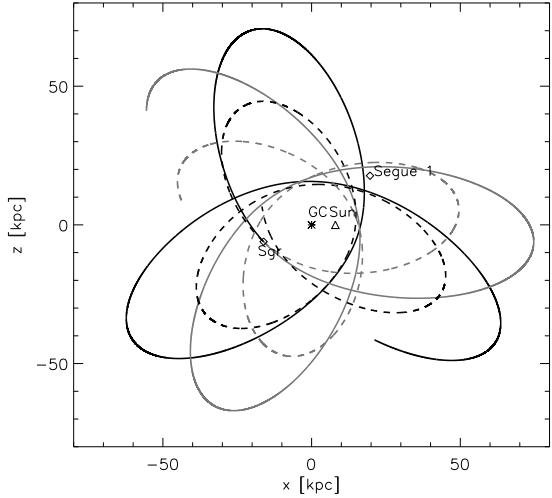}
\caption{Schematic plot of possible orbits of the Sagittarius dwarf
  in the $(x,z$) plane integrated forwards (dark grey) and backwards
  (light gray) for 2 Gyr. The position of Segue 1 is close to
  crossings of the rosette orbit. The potential is defined by
  equations (1)-(3) in Fellhauer et al. (2006). The dashed line
  corresponds to an orbit starting at right ascension $\alpha =
  283.7^\circ$, declination $\delta =-30.5^\circ$ and a heliocentric
  distance of 25 kpc (the location of the Sagittarius) with initial
  velocities $v_r=137$ kms$^{-1}$, $\mu_\alpha=-3.02$ mas yr$^{-1}$,
  $\mu_\delta = -1.49$ mas yr$^{-1}$. The solid line starts at the
  same location but has initial velocities $v_r= 130$ kms$^{-1}$,
  $\mu_\alpha = -3.44$ mas yr$^{-1}$, $\mu_\delta = -1.32$ mas
  yr$^{-1}$.}
\label{fig:rosette}
\end{figure}
\begin{figure}
\includegraphics[width=\linewidth]{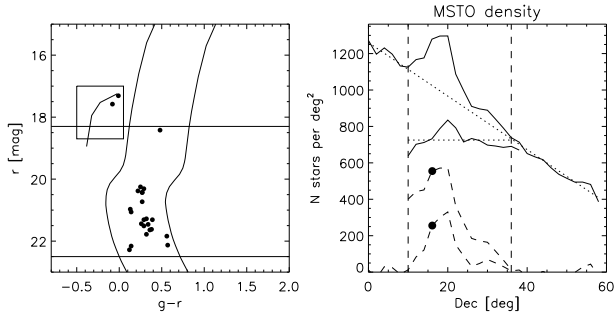}
\caption{Left: CMD showing the selection box used to pick BHB stars in
  the SDSS spectral database and the mask to pick turn-off stars for
  measuring the density of the Sagittarius stream around Segue 1.
  Possible Segue 1 members selected by \citet{Ge09} are shown as black
  dots. Right: The upper solid line shows the density profile of the
  turn-off stars selected, with the dotted line showing a linear fit
  to the background and the residuals shown as lower dashed line. We also
  estimate the background by examining stars at the same Galactic
  latitude but with the sign of longitude reversed. This results in a
  lower background estimate and a higher density for the Sagittarius
  stream (upper dashed curve). The filled circles mark the location of
  Segue 1.  The vertical dashed lines mark the boundaries of the
  Sagittarius stream.}
\label{fig:selection}
\end{figure}
\begin{figure}
\includegraphics[width=\linewidth]{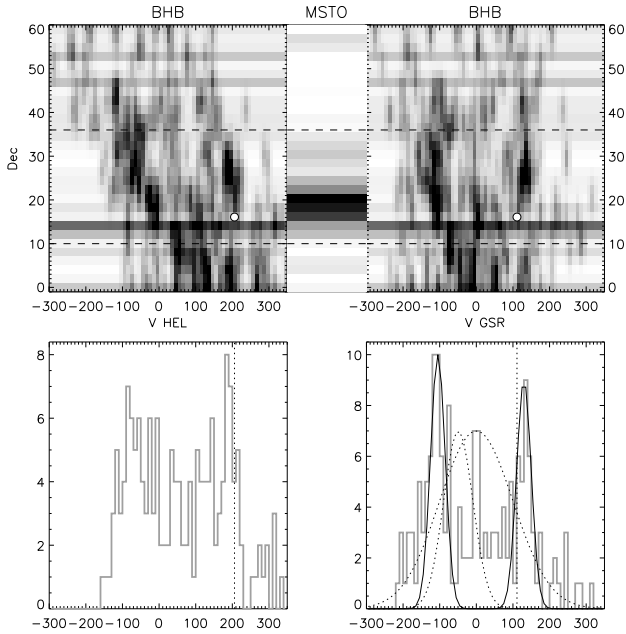}
\caption{Upper: The density of BHB stars in the plane of declination
  versus radial velocity (heliocentric for the left, Galactocentric
  for the right panel). The efficiency of the spectral follow-up is
  indicated by the horizontal bands with white representing $50 \%$
  and black representing $10 \%$. To guide the eye, the
  one-dimensional density profile of the Sagittarius stream (dashed
  curve from Fig~\ref{fig:selection}) is shown connecting the two
  panels. The dashed lines mark the boundaries of the Sagittarius
  stream.  Lower: Radial velocity distributions for stars in the
  declination range $10^\circ \le \delta \le 36^\circ$ of the
  Sagittarius stream.  The stream is most clearly identifiable in the
  right panel in which the velocities are Galactocentric. Note the two
  features at $v_{\rm GSR} = -105$ (the A and B streams) and $130$
  kms${}^{-1}$ (the C and D streams). We also show as dotted lines the
  characteristic velocity distributions of the Galactic halo and thick
  disk to emphasise that the features cannot be ascribed to these
  components. Finally, the two black distributions are Gaussian fits
  to the stream velocities.}
\label{fig:signal}
\end{figure}

\begin{figure}
\includegraphics[width=\linewidth]{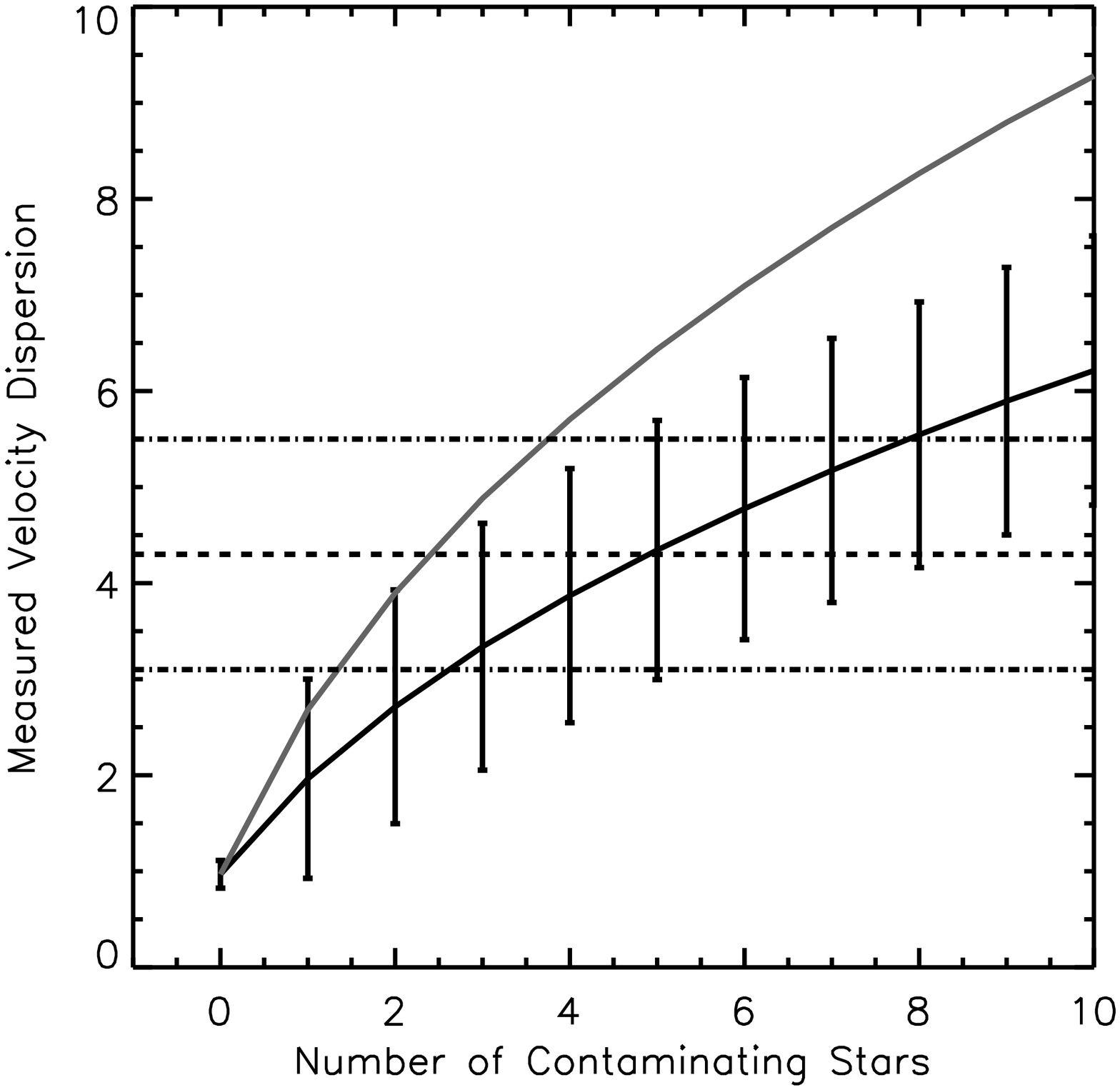}
\caption{ The measured velocity dispersion of Segue 1, as a function of
  the number of contaminating stars from the Sagittarius stream. The
  lines indicate the average of 10000 simulations, whilst the error
  bars represent the spread. The black line represents a contaminating
  population with dispersion 10 kms$^{-1}$ whilst the grey line
  represents a contaminating population with dispersion of 15
  kms$^{-1}$. The value reported by Geha et al. (2008) is shown as a
  horizontal dotted line, with the dash-dotted lines giving the
  reported error.}
\label{fig:MC}
\end{figure}

\section{Kinematics of  Segue 1}

\subsection{The Sagittarius Connection}

Segue 1 lies in a very busy area of the sky with multiple wraps of the
Sagittarius stream as well as other debris streams, as shown in the
``Field of Streams'' (Belokurov et al. 2006). It has been known since
the work of Ibata et al. (2001) that the Sagittarius tidal debris is
wrapped around the Galaxy a number of times.  There are at least 4
wraps corresponding to two streams, leading and trailing~\citep[see
e.g.,][]{Fe06}. A schematic plot is shown in Figure~\ref{fig:rosette},
in which the orbit of the Sagittarius dwarf is integrated for $\sim 2$
Gyr backwards and forwards in time. Segue 1 is located close to an
intersection of the trailing and leading arms.  The ``Field of
Streams'' traces the young leading (A) and old trailing (B) wraps,
which are closely matched in distance around the North Galactic
Cap. Simulations suggest the existence of further streams
corresponding to old leading and young trailing wraps in the same
field of view. So far, observational evidence for these wraps is
sparse.

At the location of Segue 1, the wraps have similar distances and are
composed of similar stars. The only way to identify them is through
their different kinematics. To this end, we interrogate the SDSS
spectral database to pick out blue horizontal branch stars.
Figure~\ref{fig:selection} shows the BHB branch of M92 with an
enclosed box, which is used for selection in $g-r$ versus $r$. To
eliminate false positives, we simultaneously impose the cut $0.9 < u-g
< 1.5$~\citep{Si04}. This box selects BHB stars at the approximate
distance of Segue 1.

Our rationale is to trace with BHBs a slice through the Sagittarius
stream at constant right ascension ($145^\circ < \alpha <
155^\circ$). Any features that stand out in the same range of
declination can be attributed to the Sagittarius stream.  The right
panel tells us what this range of declination is. The density of
turn-off stars is shown as the uppermost solid line, with the turn-off
stars selected via the mask shown in the left panel. The mask is wide
enough to include all stars (22, excluding 2 BHBs) in the
spectroscopically confirmed sample of \citet{Ge09}. Although the mask
is centered on the CMD of Segue 1, because of its width, it will also
pick out stars in the Sagittarius stream.

To make Sagittarius stand out more clearly, we subtract the Galactic
foreground.  The number density of Sagittarius stars depends on the
model adopted for the foreground. In one method, we take advantage of
the apparent linearity of the foreground (shown as a dotted
line). Once subtracted, this gives the lower dashed profile, which
peaks at $\sim 300$ stars per deg$^{2}$. However, at low declinations,
the fit could be in error due to a contribution from the ``Virgo
Overdensity''~\citet[see e.g.]{Be06,Ju08}. As an alternative, we
examine the density of stars at the same Galactic latitude but with
the sign of longitude reversed.  This results in a lower foreground
estimate and a higher peak density for the Sagittarius of $\sim 600$
stars per deg$^{-2}$. The Sagittarius stream is limited to
declinations satisfying $10^\circ < \delta < 36^\circ$.

Figure~\ref{fig:signal} shows the distribution of BHB stars in the
plane of radial velocity versus declination. Once the gradients due to
the projection of the Local Standard of Rest have been removed,
the two overdensities coincident with the Sagittarius stream in
declination stand out at $v_{\mathrm GSR} = -105$ kms$^{-1}$ and $130$
kms$^{-1}$. At this location, it is unlikely that these features can
be due to the Galactic thick disk or halo, as evidenced by the
velocity histograms in the lower panel. In fact, they nicely match the
velocities of the wraps of the Sagittarius stream at this location
shown in Figure~3 of Fellhauer et al. (2006). Segue 1 is in the
Sagittarius stream and appears to be moving with the same velocity as
one its wraps.  In the lower panels in Figure~\ref{fig:signal}, we
show Gaussian fits to each of the overdensities associated with
Sagittarius. The dispersions are $\sim 20$ kms$^{-1}$. Assuming
typical radial velocity errors of $\sim 15$ kms$^{-1}$ gives the
intrinsic dispersion in each wrap to be $\sim 13$ kms$^{-1}$. Of
course, this is approximate, as both measured dispersion and the
errors could be somewhat larger.

This has immediate consequences for estimates of the velocity
dispersion of Segue~1, as the problem of contamination by Sagittarius
stream stars is substantial. Let us estimate the number of Sagittarius
stars in the sample of~\citet{Ge09}. As shown in
Figure~\ref{fig:selection}, our mask already wraps around the
candidate stars of Geha et al. Hence, the total number of Sagittarius
stars at the location of Segue~1 can be read off the dashed curve in
the right-hand panel. It is between $\sim 250$ and $\sim 500$ stars
per deg${}^{2}$ depending on the model adopted for foreground
subtraction. To estimate how many of these move with $v_{\mathrm GSR}
= 130$ kms$^{-1}$, we note that the two velocity peaks in the lower
panel of Figure~\ref{fig:signal} integrate to roughly the same numbers
of stars.  So, allowing a factor of 0.5 to to account for the fact
that both leading and trailing arm stars occur at this location and
scaling to the $0.03$ deg${}^{2}$ field of view used by Geha et al, we
estimate that the final contamination in their sample is at least
$\sim 250$ stars deg${}^{-2}$ $\times 0.5 \times .03$ deg${}^{2}$,
which comes to 4 stars. This might be raised as high as 8 stars, if
the lower value for the foreground is adopted. We caution that such
estimates necessarily involve extrapolation from comparatively small
numbers of BHBs.

Could contaminating stars be responsible for the high velocity
dispersion of $4.3 \pm 1.2$ kms$^{-1}$ reported by~\citet{Ge09}?
Figure~\ref{fig:MC} shows the results of Monte Carlo simulations to
gauge the importance of this contamination. The Sagittarius stream
stars are assumed to have a velocity dispersion of $10$ kms$^{-1}$ and
Segue~1 stars to have a velocity dispersion of $1$ kms$^{-1}$ (typical
for a globular cluster). In the simulations, samples of 24 stars are
generated, with the number of contaminating Sagittarius stream stars
allowed to vary between 0 and 10.  The simulations are run for 10000
iterations.  The points on the plot are the average dispersion
measured in the 10000 samples of 24 stars and the error bars represent
the spread.

The number of contaminating Sagittarius stars required to inflate the
velocity dispersion from its true value (in the simulation) of $1$
kms$^{-1}$ to the value reported by Geha et al is surprisingly small
-- perhaps even one contaminating star is enough! Recall that we
estimate the actual number of contaminants in the Geha et al sample is
probably between 4 and 8. Although the kinematic selection of Geha et
al is sufficient to exclude stars from one of the wraps of the
Sagittarius, it is not sufficient to exclude stars from the wrap
moving with $v_{\mathrm GSR} = 130$ kms$^{-1}$. A very modest
contamination from this wrap can cause an anomalously high velocity
dispersion to be reported.

\begin{figure*}
	\subfigure[]{
	\includegraphics[width=0.38 \linewidth]{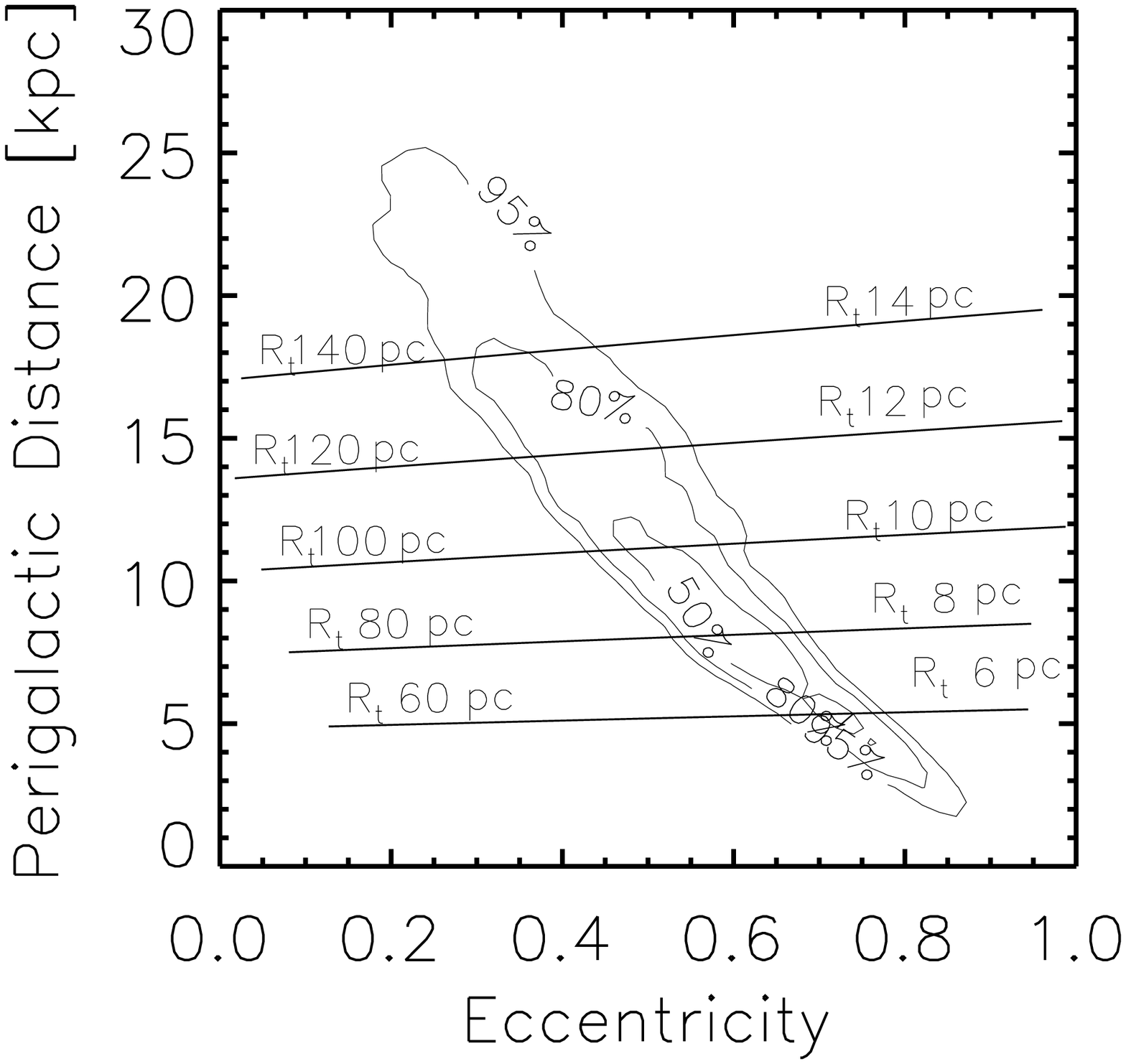}}
	\subfigure[]{
	\includegraphics[width=0.38 \linewidth]{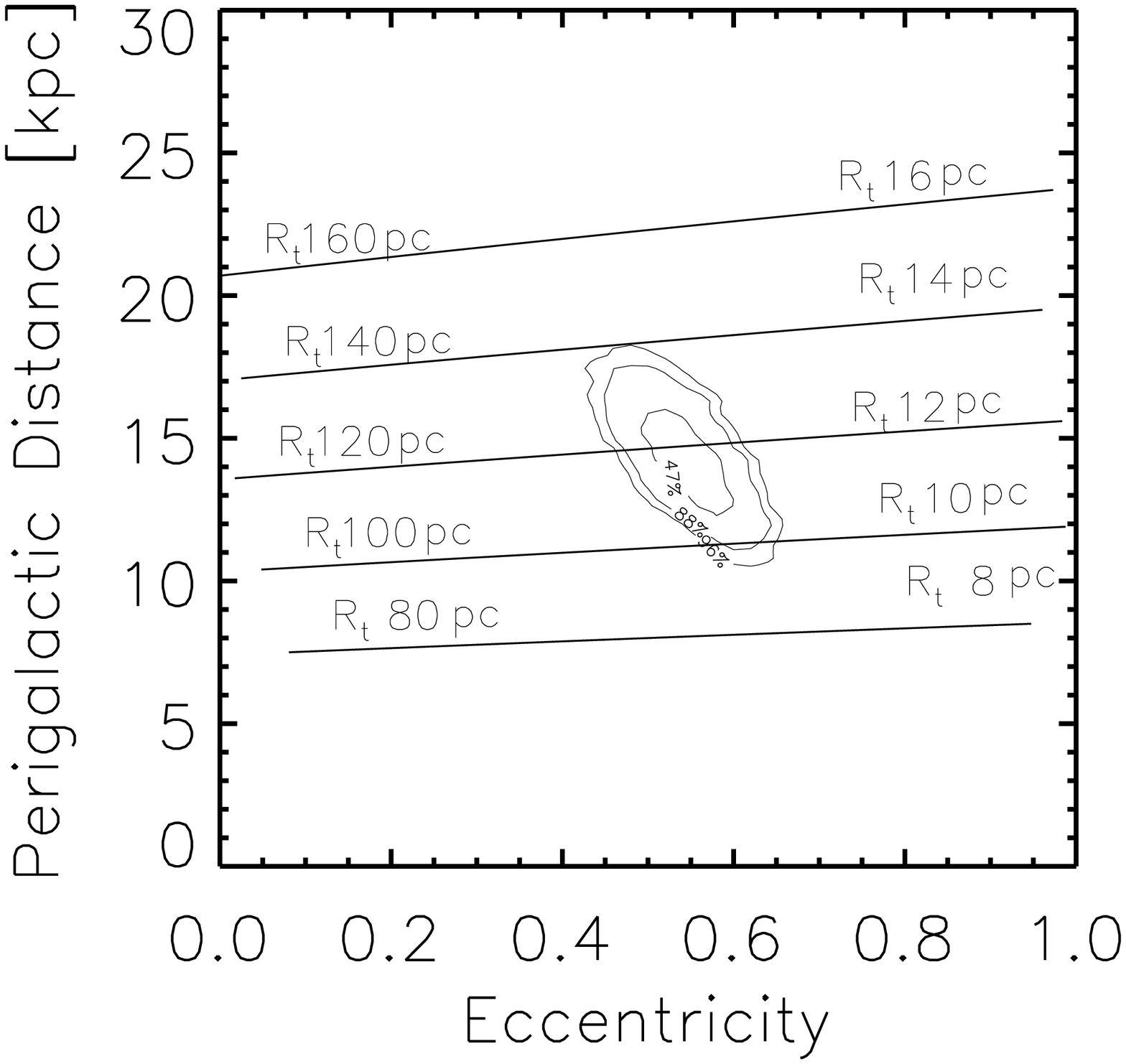}}
      \caption{Left: The results of orbit integration with initial
        velocity $v_r \approx 210$ kms$^{-1}$ and a spread of
        tangential velocities ($75$ kms$^{-1}$ -- $225$ kms$^{-1}$ and
        $m=1000 M_{\sun}$ (left contour labels) and $10^6M_{\sun}$
        (right contour labels). The tidal radii implied by these
        orbits suggest that if Segue 1 was a globular cluster it
        should show signs of tidal disruption. Right: The same, but
        under the assumption that Segue 1 is following the Sagittarius
        orbit of Fellhauer et al (2006).}
	\label{fig:orbit_info}
\end{figure*} 

\begin{figure*}
\includegraphics[width=0.76\linewidth]{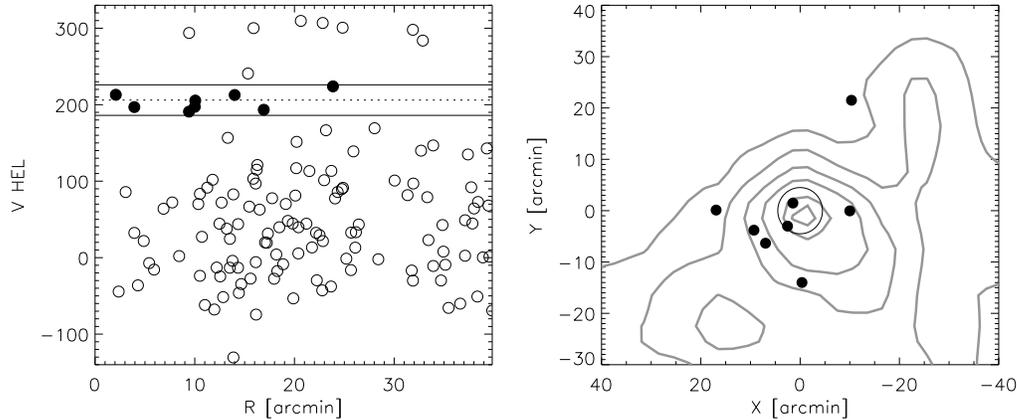}
\caption{Left: Heliocentric radial velocity of giant stars from the
  AAT survey plotted against angular distance from the center of Segue
  1. The filled circles show 8 stars that have the same kinematics as
  Segue 1 ($206 \pm 20$ kms$^{-1}$). Right: The locations of the 8
  stars are spread over a wide field of view. The grey contour lines
  show the density distribution around Segue 1 as taken from the
  middle panel of Figure~\ref{fig:optfilters1}. The black circle marks
  the half-light radius of Segue 1. The AAT candidates extend out to
  $\sim 5$ half-light radii.}
\label{fig:AAT}
\end{figure*}

\subsection{Tidal Influence}

Assuming that Segue 1 is on a Sagittarius-like orbit, its radial and
tangential velocities are sufficiently well constrained to ask what
effect Galactic tides would have on the
object. Figure~\ref{fig:orbit_info} shows the results of orbit
integrations, in which a test particle is placed at the location of
Segue 1 with a variety of initial velocities.  The particle travels
through the Milky Way potential as defined in Fellhauer et al. (2006)
for roughly 2 periods. We then measure the perigalactic distance of
the orbit as well as its eccentricity. The contours show the regions
in which the points lie in the plane of perigalactic distance and
eccentricity. We then calculate the tidal radius for a cluster of mass
$m$ travelling on such an orbit using~\citet{In83}
\begin{equation}
r_{\rm tidal} = \frac{2}{3} \left(\frac{m}{(3+e)
  M_{\rm P}}\right)^{\frac{1}{3}} R_{\rm P}.
\end{equation}
Here, $R_P$ is the perigalactic distance, $e$ is the eccentricity, and
$M_P$ is the mass of the galaxy interior to $R_P$. In both panels of
the figure, the radial velocity is $v_r \approx 210$ kms$^{-1}$
(heliocentric) or 118 kms$^{-1}$ (Galactocentric). But, panel (a)
shows the results with initial tangential velocities corresponding to
a typical Galactic range ($75$ kms$^{-1}$ to $225$ kms$^{-1}$), and
with cluster masses of $m=1000 M_{\sun}$ and m=$10^6 M_{\sun}$. Panel
(b) assumes the tangential velocity is chosen to correspond
approximately to that of the old leading arm 215 kms$^{-1}$
Galactocentric).

From both panels, we find that if Segue 1 is a globular cluster with
mass around $1000 M_{\sun}$, then its tidal radius would be smaller
than its observed half-light radius which implies that it could not
survive on such an orbit for long and would be in the throes of
destruction.  If Segue 1 is a dwarf galaxy with mass of $10^6 M_{\sun}$,
the estimated tidal radius is much larger than the measured half-light
radius and we would not expect to see any signs of tidal disruption

\section{Discussion and Conclusions}

The distinction between star clusters and dwarf galaxies is largely
based on size. There is a factor of $\gtrsim 10$ difference in the
characteristic sizes of clusters and galaxies. Despite differences in
the velocity dispersion, this translates into an order of magnitude
difference in the estimated mass-to-light ratios, under the assumption
of virial equilibrium.

The example of Pal 5 shows the dangers of assuming virial
equilibrium. Pal 5 has a half-light radius of $\sim 20$ pc, one of the
largest of the Milky Way globular clusters. It is also well-known to
be disintegrating under the Galactic tidal field, and so its
half-light radius exceeds its tidal radius. The most distant stars are
therefore already unbound and belong properly speaking to Pal 5's
tidal tails.  The assumption of virial equilibrium would lead to a
mass-to-light ratio that is in serious error. As noted by Dehnen
(2004), the final disruption of Pal 5 happens very quickly, and most
mass is lost during the final few percent of its lifetime.  Another
example of a dissolving Milky Way satellite is provided by Ursa Major
II (UMa~II). In the model of Fellhauer et al. (2007), the orbit of
UMa~II has pericentric and apocentric distance similar to that of the
Sagittarius. The fate of UMa~II is very similar to that of Pal 5, but
the final disruption happens on a longer timescale, comparable to the
orbital period of 1 Gyr (see Figure 6 of Fellhauer et
al. 2007). Suppose we were to observe Pal 5 or UMa~II at a future time
equal to a fraction of their orbital period. Then, just a few faint
stars will mark the nucleus, whilst any streams will become blurred
and of lower contrast as they diffuse in the Milky Way potential.  In
other words, we will see something like Segue 1.

Of course, to prove the hypothesis of extra-tidal stars for Segue 1,
what is needed is wide-field kinematical data. At the moment, the only
published study is by \cite{Ge09}. They measured the radial velocities
of 24 stars and concluded that Segue 1 has a heliocentric velocity of
$\sim 206$ kms${}^{-1}$ and a velocity dispersion of $4.2$
kms$^{-1}$. The systemic velocity therefore coincides with both the
predicted velocity of the Sagittarius stream~\citep{Fe06} and the
measured velocity as recorded in this paper. This not only strengthens
the argument that Segue 1 is a star cluster, originating from the
Sagittarius, but also raises the question of the levels of
contamination in kinematically selected datasets around Segue 1's
location.  In particular, such contaminants may be inflating the
velocity dispersion, and hence the inferred mass-to-light ratio, of
Segue~1.

Taking advantage of the wide field of view of the Anglo-Australian
Telescope/ AAOmega fiber-fed spectrograph combination~\footnote{See
  http://www.aao.gov.au/local/www/aaomega/}, Gilmore and collaborators
have observed a $0.8^\circ \times 0.8^\circ$ area around Segue~1.
Observing parameters were the same as those for the related study of
Bootes~1 (see Norris et al. 2008).  Targets are selected from the SDSS
photometric data to be broadly consistent with the Segue 1 CMD
published by Belokurov et al. (2007) and a radial velocity accuracy of
$<10$ kms$^{-1}$ is attained. Analysis of this dataset is ongoing and
will be presented elsewhere. Here, we are primarily concerned with
existence of extratidal stars. The left panel of Figure~\ref{fig:AAT}
shows the dataset plotted in the plane of heliocentric radial velocity
and angular distance from Segue 1. Applying a kinematic cut of $206
\pm 20$ kms$^{-1}$ picks out the 8 likely members, which are
reasonably well separated from the Galactic foreground
populations. The right panel shows the location of the likely members
as compared to the density distribution inferred from the optimal
filter. Not only do the candidates extend well beyond the half-light
radius marked by the thin circle, but they also follow in broad
outline the distribution suggested by the photometric analysis.  This
confirms the wide spatial distribution of stars with kinematics like
Segue 1 in this field.

Accordingly, we favour the scenario in which Segue 1 is a star cluster
stripped early on from the Sagittarius galaxy. This provides a natural
explanation for the fact that all four of the phase space coordinates
that are measured coincide with those of the Sagittarius stream.  What
is needed to make the argument complete is a demonstration that the
metallicity of Segue 1 is compatible with an origin in the Sagittarius
galaxy. \citet{Ge09} measured the metallicity of one star as [Fe/H] =
-3.3, and the metallicity of a further 13 as [Fe/H] = -1.8. The latter
is consistent with the metallicity of Sagittarius.  The former is very
surprising, as there are no Sagittarius clusters with a metallicity
lower than -2.2. However, the declination and radial velocity of Segue
1 place it in the very oldest wrap of material, and so it must have
been stripped first and have resided in the very outskirts of the
Sagittarius progenitor. This could explain variations in the
metallicity of Segue 1 and other parts of the Sagittarius stream.
Regardless of whether the metal-poor star of \citep{Ge09} is a member
of Segue 1 or not, it signifies the detection of a substantial
metallicity gradient in the Sagittarius progenitor, although this is
more extreme than has been measured so far~\citep{Ch07}. In any case,
the claimed status of Segue 1 as the least luminous galaxy is very
uncertain.

\section*{Acknowledgments} 
We thank the referee, Rodrigo Ibata, for a careful reading of the
paper.  MNO is funded by the Gates Cambridge Trust, the Isaac Newton
Studentship fund and the Science and Technology Facilities Council
(STFC), whilst VB acknowledges financial support from the Royal
Society.

Funding for the SDSS and SDSS-II has been provided by the Alfred P.
Sloan Foundation, the Participating Institutions, the National Science
Foundation, the U.S. Department of Energy, the National Aeronautics
and Space Administration, the Japanese Monbukagakusho, the Max Planck
Society, and the Higher Education Funding Council for England. The
SDSS Web Site is http://www.sdss.org/.

\label{lastpage}


\begin{thebibliography}{99}

\bibitem[Adelman-McCarthy et al.(2008)]{Ad08} 
Adelman-McCarthy, J.~K., et al.\ 2008, ApJS, 175, 297 

\bibitem[\protect\citeauthoryear{Belokurov et
    al.}{2006}]{Be06} Belokurov, V., et al., 2006, ApJ, 642, L137

\bibitem[\protect\citeauthoryear{Belokurov et al.}{2007}]{Be07}
  Belokurov, V., et al. 2007, ApJ, 654, 2

\bibitem[Carollo et al.(2007)]{Ca07} Carollo, D., et al.\ 2007, Nature,
  450, 1020

\bibitem[Chou et al.(2007)]{Ch07} Chou, M.-Y., et al.\ 2007, 
ApJ, 670, 346 

\bibitem[\protect\citeauthoryear{Clem}{2005}]{Cl05} Clem, J. L.,
  2005, PhD Thesis, University of Victoria

\bibitem[Dehnen et al.(2004)]{2004AJ....127.2753D} Dehnen, W., Odenkirchen, 
M., Grebel, E.~K., \& Rix, H.-W.\ 2004, AJ, 127, 2753 

\bibitem[Fellhauer et al.(2006)]{Fe06} Fellhauer, M., et 
al.\ 2006, ApJ, 651, 167 

\bibitem[Fellhauer et al.(2007)]{Fe07} Fellhauer, M., et 
al.\ 2007, MNRAS, 375, 1171 

\bibitem[Fukugita et al.(1996)]{Fu96} Fukugita, M., Ichikawa, T.,
  Gunn, J.~E., Doi, M., Shimasaku, K., \& Schneider, D.~P.\ 1996, AJ,
  111, 1748

\bibitem[Geha et al.(2009)]{Ge09} Geha, M., Willman,
  B., Simon, J.~D., Strigari, L.~E., Kirby, E.~N., Law, D.~R., \&
  Strader, J.\ 2009, ApJ, 692, 1464

\bibitem[Gunn et al.(1998)]{Gu98} Gunn, J.E. et al. 1998, AJ, 116,
  3040
 
\bibitem[Gunn et al.(2006)]{Gu06} Gunn, J.E. et al. 2006, ApJ, 131, 2332

\bibitem[Hogg et al.(2001)]{Ho01} Hogg, D.W., Finkbeiner, D.P.,
Schlegel, D.J., Gunn, J.E. 2001, AJ, 122, 2129

\bibitem[Ibata et al.(2001)]{2001ApJ...551..294I} Ibata, R., Lewis,
  G.~F., Irwin, M., Totten, E., \& Quinn, T.\ 2001, ApJ, 551, 294
 
\bibitem[Innanen et al.(1983)]{In83} Innanen, K.~A., Harris, 
W.~E., \& Webbink, R.~F.\ 1983, AJ, 88, 338 

\bibitem[Ivezi\'{c} et al.(2004)]{Iv04} Ivezi\'{c}, \v{Z}. et al.\
  2004, AN, 325, 583

\bibitem[Juri{\'c} et al.(2008)]{Ju08} Juri{\'c}, M., et 
al.\ 2008, ApJ, 673, 864 

\bibitem[Martin et al.(2008)]{Ma08} Martin, N.~F., de Jong, 
J.~T.~A., \& Rix, H.-W.\ 2008, ApJ, 684, 1075 

\bibitem[Norris et al.(2008)]{No08} Norris, J.~E.,
  Gilmore, G., Wyse, R.~F.~G., Wilkinson, M.~I., Belokurov, V., Evans,
  N.~W., \& Zucker, D.~B.\ 2008, ApJ, 689, L113

\bibitem[Odenkirchen et al.(2002)]{Od02} Odenkirchen, M., Grebel,
  E.~K., Dehnen, W., Rix, H.-W., Cudworth, K.~M.\ 2002, AJ, 124, 1497

\bibitem[\protect\citeauthoryear{Odenkirchen et al.}{2003}]{Od03}
  Odenkirchen, M., et al. 2003, AJ, 126

\bibitem[Oke \& Gunn(1983)]{1983ApJ...266..713O} Oke J.~B., Gunn
  J.~E. 1983, ApJ, 266, 713                                     

\bibitem[\protect\citeauthoryear{Padmanabhan et
    al.}{2007}]{Pa07} Padmanabhan, N., et al. 2007,
  arXiv:astro-ph/0703454
                                           
\bibitem[Pier et al.(2003)]{Pi03} Pier, J.R., Munn, J.A., Hindsley,
  R.B, Hennessy, G.S., Kent, S.M., Lupton, R.H., Ivezic, Z. 2003, AJ,
  125, 1559

\bibitem[\protect\citeauthoryear{Schlegel, Finkbeiner, \&
    Davis}{1998}]{Sc98} Schlegel, D.J., Finkbeiner, D.P., \&
  Davis, M., 1998, ApJ, 500

\bibitem[Sirko et al.(2004)]{Si04} Sirko, E., et al.\ 2004, 
AJ, 127, 899 

\bibitem[Smith et al.(2002)]{Sm02} Smith, J.~A., et al.\
2002, AJ, 123, 2121

\bibitem[Tucker et al.(2006)]{2006AN....327..821T} Tucker, D.~L., et
  al.\ 2006, AN, 327, 821

\bibitem[\protect\citeauthoryear{Willman et al.}{2005}]{Wi05} Willman,
  B., et al. 2005, ApJ, 629, L85

\bibitem[\protect\citeauthoryear{Walsh et al.}{2007}]{Wa07} Walsh S.,
  Jerjen H., Willman B. 2007, ApJ, 629, L85

\bibitem[\protect\citeauthoryear{Walsh et al.}{2008}]{Wa08} Walsh S.,
  Willman B., Sand D., Harris J., Seth A., Zaritsky D., Jerjen H. 2008, ApJ, 688, 245

\bibitem[\protect\citeauthoryear{York et al.}{2007}]{Yo00} York D., et
  al. 2000, AJ, 120, 1579


\end{thebibliography}
\end{document}